\documentclass[superscriptaddress,
 reprint,
 amsmath,
 amssymb,
 aps
]{revtex4-2}

\usepackage{graphicx}% Include figure files
\usepackage{dcolumn}% Align table columns on decimal point
\usepackage{bm}% bold math
\usepackage[english]{babel}

\usepackage[a4paper,top=2cm,bottom=2cm,left=1.8cm,right=1.8cm]{geometry}

\usepackage{amsmath}
\usepackage{graphicx} %, subcaption}
\usepackage[colorlinks=true, allcolors=blue]{hyperref}
\usepackage{braket} % for \ket, \bra, \braket, etc.
\usepackage{tikz}
\usepackage{quantikz}
\usetikzlibrary{quantikz2}
\usepackage{xcolor}
\usepackage{amsmath,amssymb}
\usepackage{placeins}
\usepackage{hyperref}
\usepackage{multirow}
\usepackage[caption=false]{subfig}
\usepackage{stackengine}

\usepackage[status=final]{fixme} % hide all comments

\usepackage[normalem]{ulem} % Used for \sout{} strikethrough

\begin{document}

\preprint{APS/123-QED}

\title{Blueprint for a fault-tolerant compound photon--atom quantum architecture}

\author{Geva Arwas}
\author{Doron Azoury}
\author{Daniel Azses}
\author{Orel Bechler}
\author{Dana Ben Porath}
\affiliation{Quantum Source Labs, Ness Ziona 74036, Israel}

\author{Barak Dayan}
\affiliation{Quantum Source Labs, Ness Ziona 74036, Israel}
\affiliation{AMOS and Department of Chemical Physics, Weizmann Institute of Science, Rehovot 76100, Israel}

\author{David Dentelski}
\author{Yaron Jarach}
\author{Nadav Kandel}
\author{Aviad Landau}
\author{Yair Margalit}
\affiliation{Quantum Source Labs, Ness Ziona 74036, Israel}

\author{Alexander Poddubny}
\affiliation{Quantum Source Labs, Ness Ziona 74036, Israel}
\affiliation{Department of Physics of Complex Systems, Weizmann Institute of Science, Rehovot 76100, Israel}

\author{Michael Slutsky}
\email{michael@qs-labs.com}
\author{Konstantin Yavilberg}
\affiliation{Quantum Source Labs, Ness Ziona 74036, Israel}

\date{\today}

\begin{abstract}

Fault-tolerant quantum computing requires architectures that simultaneously address scalability, connectivity, and error correction under realistic noise constraints. We present a compound photonic--atomic quantum computing platform that uses cavity quantum electrodynamics (cavity QED) to realize near-deterministic entangling operations between flying photonic qubits and stationary atomic qubits. Photons enable long-range connectivity and massive scalability through measurement-based quantum computing (MBQC), whereas atoms provide reusable, near-deterministic resources for photon generation and entanglement.  This combined approach solves the vast inefficiency of purely photonic quantum computing platforms, which rely on highly probabilistic operations.

The central primitive is a symmetrized version of the Duan--Kimble photon--atom controlled-phase (CZ) gate, which is robust against experimental imperfections and facilitates high fidelity. We present a comprehensive platform based on single $^{87}$Rb atoms coupled to optical cavities, including protocols for state preparation, measurement, photon generation, and entangling operations on timescales of tens of nanoseconds. At the architectural level, we show how large-scale cluster states suitable for MBQC can be generated efficiently with effectively unrestricted qubit connectivity and reduced resource overhead through atomic reuse.

We analyze fault-tolerant operation based on the Raussendorf--Harrington--Goyal (RHG) lattice and introduce cluster-state generation schemes tailored to the compound hardware. A hardware-aware noise model captures asymmetric loss processes and correlated errors between the photonic and atomic subsystems. Logical memory simulations yield a photon-loss threshold of approximately $2.6\%$ per physical gate, corresponding to $\sim15\%$ total loss over a photon trajectory. Beyond memory, we demonstrate that the full Clifford gate set---logical Hadamard, phase, and CNOT---can be implemented transversally or fold-transversally within the RHG framework, achieving thresholds matching the value for the identity channel. We further propose two complementary routes to non-Clifford resource-state preparation, based on code teleportation and magic state cultivation, both reformulated within the foliated cluster-state architecture.

\end{abstract}

\maketitle
\tableofcontents

\section{Introduction}

\subsection{Background}

Utility-scale fault-tolerant quantum computers are expected to require millions of physical qubits in order to bridge the “quantum chasm”~\cite{chasm, eisert2026mindgapsfraughtroad}
between the error rates of $10^{-3}$--$10^{-4}$ of current state-of-the-art qubits and gates, to the  $< 10^{-12}$ error rates required for useful quantum applications~\cite{gidney2025factor2048bitrsa, eisert2026mindgapsfraughtroad}. 
This necessitates encoding quantum information into fault-tolerant logical elements, using $10^2-10^3$ physical qubits per logical element. Accordingly, algorithms involving a few thousand high-quality logical qubits are expected to involve $\sim10^6$ physical qubits on realistically noisy hardware~\cite{gidney2025factor2048bitrsa, eisert2026mindgapsfraughtroad}. Equally vital to the fault-tolerant operation is the connectivity, namely the ability to perform entangling gates between distant physical qubits~\cite{Cohen_2022}.
High connectivity facilitates efficient quantum error-correction codes and fault-tolerant logical gates, and provides greater flexibility in circuit compilation by removing spatial constraints on logical two-qubit gates~\cite{Cohen_2022, Baspin_2022, filippov2026architectingdistributedquantumcomputers}.

Modern matter-based quantum computing platforms encounter substantial barriers to scalability for technical and fundamental reasons. Superconducting-qubit platforms are challenged by the need to overcome the coupling constraints of planar architectures and go beyond 
nearest-neighbor interactions~\cite{mohseni2024build,megrant2025scaling,wang2026demonstration,bravyi2024high,mathews2026placing}. 
In trapped-ion systems,  motional-mode crowding and degraded gate performance constitute a major hurdle for entangling operations between distant qubits~\cite{landsman2019pra,johnson2025nonlinear}. 
Neutral-atom platforms offer reconfigurable connectivity through physical qubit shuttling, but at a substantial cost in clock speed. Shuttling operations and mid-circuit measurements result in quantum error correction (QEC) round times of several milliseconds---roughly three orders of magnitude slower than in superconducting platforms---and introduce a logical-depth overhead that grows with transport distance~\cite{evered2023high,reichardt2024fault,khan2026architecting}. In addition, the number of matter-based qubits that can be maintained in a single controlled environment is limited. Scaling to millions of qubits therefore requires modular architectures, where reliable inter-module connections become an additional major challenge~\cite{knorzer2025distributed}.

Photonic qubits, and specifically dual-rail photonic qubits --- encoded, for example, in path or polarization degrees of freedom --- circumvent many of the scaling limitations inherent to matter-based platforms ~\cite{abughanem2026toward,zhu2026recent,bourassa2021blueprint}.
As photons do not interact
with one another and couple only weakly to their environment, photonic qubits exhibit
exceptionally low decoherence, requiring neither cryogenic temperatures nor ultra-high
vacuum. Single-qubit gates can be implemented with high fidelity using linear optical
components~\cite{kok2007linear,rudolph2017optimistic,romero2024photonic},
and optical fiber delay lines serve as near-noise-free quantum
memories~\cite{bombin2021interleaving}. Once inter-qubit entanglement is established,
qubit connectivity in photonic architectures is effectively unrestricted, considerably
clarifying the path toward fault-tolerant quantum
computation, using measurement-based (MBQC) and fusion-based approaches ~\cite{raussendorf2001one,raussendorf2006fault,raussendorf2007topological,bartolucci2023fusion,briegel2009measurement}.
However, attaining this inter-qubit entanglement, namely the construction of the required entangled photonic
states (known as cluster or graph
states~\cite{briegel2001persistent, hein2004multiparty}) is the main bottleneck of the photonic approach. 
Without the mediation of single quantum emitters, single photons cannot be deterministically generated or entangled, as linear optics operations and even Kerr-type interactions cannot provide the required  nonlinearity~\cite{PhysRevA.73.062305,PhysRevLett.133.113601}.
Single-photon sources based on parametric photon-pair
generation~\cite{burnham1970observation,kwiat1999ultrabright,pittman2002single,%
sharping2006generation,reimer2014integrated,silverstone2014on} are inherently probabilistic,
as are linear-optics based entangling gates 
~\cite{browne_rudolph2005resource,kok2007linear,varnava2008how,li2015resource,bartolucci2021creation}. 
Furthermore, high-fidelity entangling operations require photons from independent sources to be mutually indistinguishable, imposing stringent demands on source quality~\cite{chan2025practical,pollet2026near}.  
Overall, scaling such platforms to utility-scale operation requires heralded protocols with active multiplexing, incurring an overhead of roughly six orders of magnitude in source and gate resources~\cite{li2015resource,bartolucci2025comparisonschemeshighlyloss}.

\begin{figure*}\centering
\includegraphics[width=0.9\linewidth]{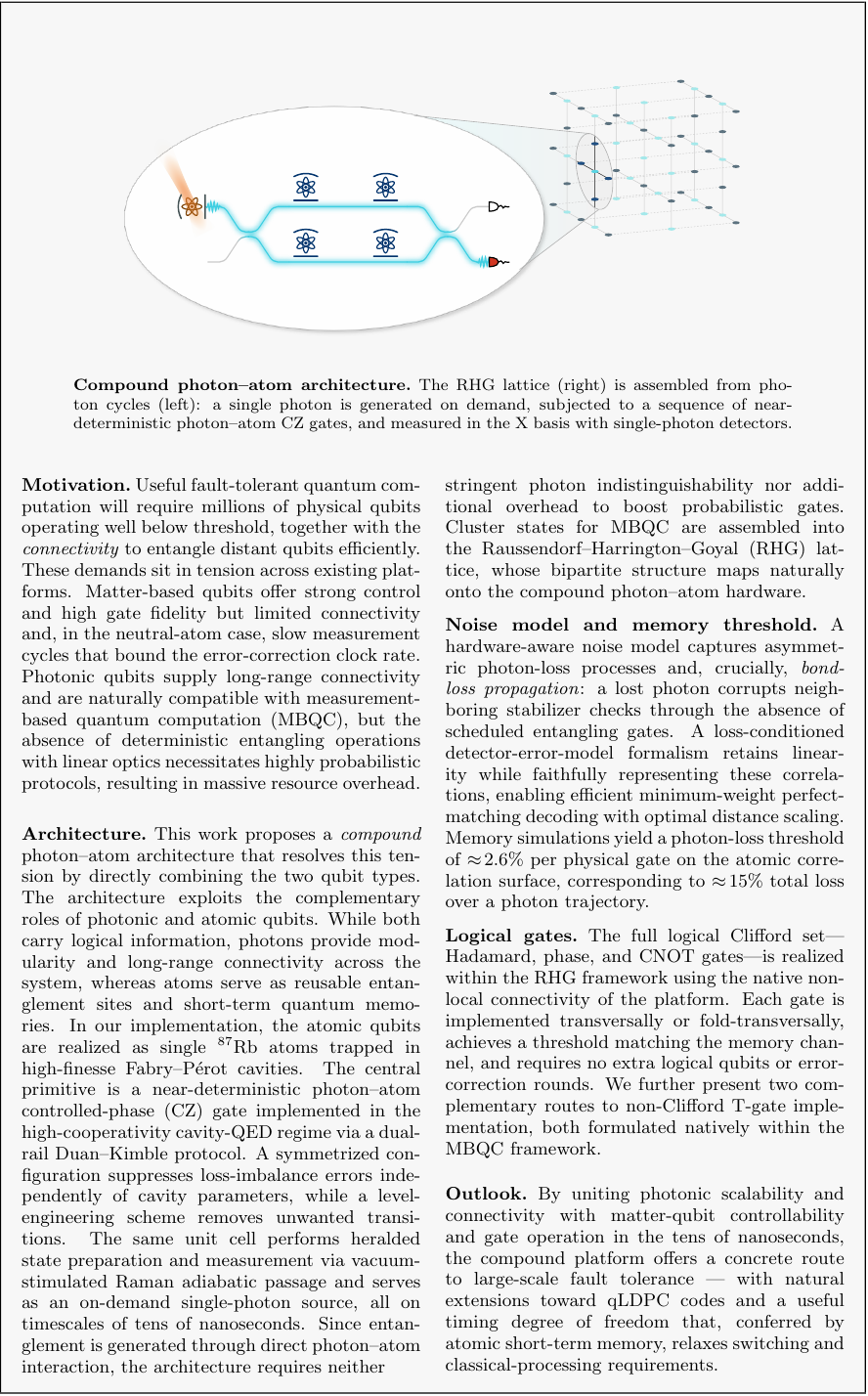}\\
\noindent\textbf{Box 1.} Compound photon--atom architecture overview. 
\end{figure*}

\subsection{The proposed solution, or why yet another modality?}

Our compound photon--atom architecture, summarized in Box~1, provides large-scale deterministic generation and entanglement of photonic qubits through an atom-mediated gate-based approach~\cite{PRXQuantum.6.010340}.
The core unit of our system is a single $^{87}$Rb atom trapped within the mode of a high-finesse optical cavity tuned near the $D_1$ transition. Harnessing high-finesse small mode-volume microresonators (see details in Sec.~\ref{subsec: Physical layer - Unit-Cell}), the system operates in the high-cooperativity regime, where appropriately prepared atomic and photonic states can interact with near-unit probability \cite{PhysRevLett.68.1132,haroche2013exploring}. In particular, a high-fidelity symmetric CZ gate is implemented using a modified Duan--Kimble protocol~\cite{duan_kimble_cz, reiserer2014quantum,PRXQuantum.6.010340}, assisted by external laser control fields, as detailed in Sec.~\ref{sec:entangling-gate}.

In the next section, we present in detail the principal quantum protocols required to construct the building blocks of fault-tolerant quantum computation from these core units. For now, we simply assume that these building blocks are already in place and focus instead on the high-level architectural features that make this system a promising approach to fault-tolerant quantum computation.

In this architecture, a large fraction of the physical qubits are photonic. As in fully photonic approaches, computation is naturally carried out within the MBQC framework. However, near-deterministic photon--atom CZ gates allow cluster-state generation without the vast resource overhead required with probabilistic linear-optics entangling operations. In addition, the same atom-cavity system can serve as a near-deterministic single-photon source using vacuum-stimulated Raman adiabatic passage (vSTIRAP) \cite{vitanov2017stimulated,hennrich2000vacuum,kuhn2002deterministic,kuhn2010cavity, reiserer2015cavity}, avoiding the low ($\sim10^{-2}$) efficiency associated with heralded probabilistic single photon generation protocols.
Furthermore, this architecture provides effectively unrestricted qubit connectivity, enabling flexible implementations of advanced quantum error-correction codes, such as quantum low-density parity-check (qLDPC) codes~\cite{vasic2025quantum,panteleev2022asymptotically,leverrier2022quantum,lin2022good,bravyi2024high}, as well as nonlocal logical gates and optimized fault-tolerant circuit compilations. 
Moreover, the fact that the atomic qubits can be reset and reused after measurement allows efficient resource optimization that minimizes the number of required cavities and the associated optical-control infrastructure by more than an order of magnitude, as discussed in Sec.~\ref{sec:GS}.

Finally, a major advantage over neutral-atom and trapped-ion architectures is the ability to perform cavity-enhanced entangling gates and state-preparation and measurement (SPAM) operations on $\sim 10$~ns timescales, yielding orders-of-magnitude speedups (especially in measurement), placing this architecture in a regime comparable to superconducting qubits~\cite{Jiang2025}. 
At the same time, the long coherence time of the atoms allows using them as dynamic memories, providing this compound architecture with a useful temporal degree of freedom, and accordingly relaxing requirements on control and classical processing.

\section{Physical layer}
\subsection{The unit cell}
\label{subsec: Physical layer - Unit-Cell}
\begin{figure*}
    \centering
    \includegraphics[width=1.0\linewidth]{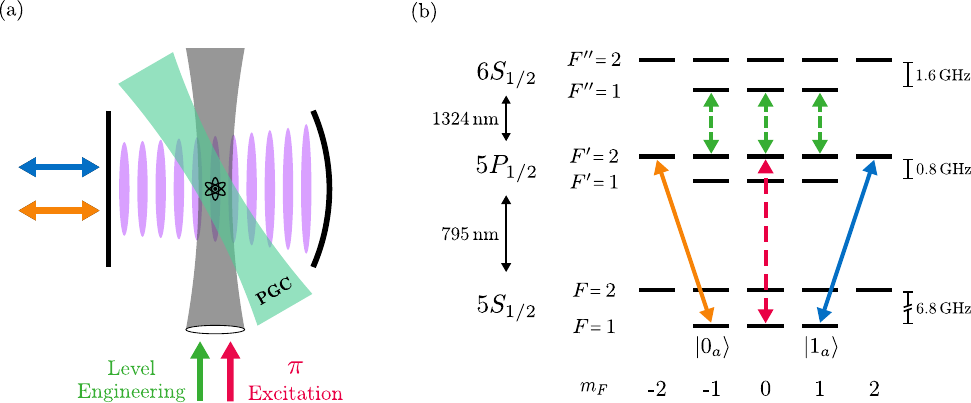}
    \caption{(a) The unit cell structure. A 1D PGC beam (light-green) loads and cools the atom into the optical lattice dipole trap (purple). The flat mirror is used for input-output coupling. Additional beams, such as $\pi$-polarized (red and green) are used for several protocols and level engineering. 
    (b) $^{87}\mathrm{Rb}$ atom level structure. The general structure of the $D_1$ line $(5S_{1/2}\leftrightarrow 5P_{1/2})$ is depicted. The hyperfine splitting of the ground and excited states are shown, together with the allowed dipole transitions, driven by $\pi$ (red dashed arrow), and $\sigma_\pm$ polarized light (blue/orange arrows). The atomic qubit space $\lvert F = 1, m_F = \pm 1\rangle$ in the ground state is also shown. The green arrows indicate the laser-induced coupling between the $5P_{1/2}$ and the higher $6S_{1/2}$ states utilized for the level-engineering process.}
    \label{fig:UC-level-structure}
\end{figure*}

The unit cell is the elementary cavity-QED module of the architecture. It consists of a single $^{87}\mathrm{Rb}$ atom trapped in an optical resonator, as shown in Fig.~\ref{fig:UC-level-structure}(a). The cell is complemented by a set of optical control fields required for coherent manipulation,
atom trapping, and cooling. The unit cell acts as a reusable node for all principal operations of the platform, including atomic state preparation and measurement, single-photon generation, and photon--atom entangling gates.

The cavity is a single‑sided Fabry--Pérot resonator, formed by two opposing micro-mirrors \cite{hunger2010fiber, pfeifer2022achievements}. One mirror is flat and serves as the input--output coupler, while the second mirror is curved with a radius of curvature of $R\sim \mathrm{30~\mu m}$. The cavity is locked to the rubidium $D_1$ transition at $795\,\mathrm{nm}$, with a cavity length of approximately $L\sim~\mathrm{25~\mu m}$. The cavity supports the fundamental TEM$_{00}$ spatial mode with two degenerate orthogonal polarization eigenmodes. In the paraxial approximation, the spatial dependence of the mode amplitude is given by
\begin{equation}
\label{eq: cavity field formula}
E(r,z) \propto \frac{e^{-r^2/w^2(z)}}{w(z)}
\sin\!\left(kz+\frac{k r^2}{2R(z)}\right),
\end{equation}
where $w(z)$ and $R(z)$ are the beam radius and wavefront radius of curvature of the mode at $z$, and $k$ is the mode wave number. 
By choosing the input-output mirror to be flat we enable near-unity mode overlap between the incoming mode and the mode of the cavity, a crucial parameter in Fabry--Pérot micro-cavities~\cite{gallego2016high}.

The decay of the cavity field in both modes is characterized by the rate $\kappa = \kappa_e + \kappa_i$, where $\kappa_e$ denotes the cavity input-output coupling, and $\kappa_i$ captures the intrinsic loss mechanisms, specifically the scattering and absorption effects of the mirrors.  
Assuming realistic mirror coating performance \cite{pfeifer2022achievements}, we can achieve values as low as $\kappa_i/2\pi \approx 3\,$MHz. This corresponds to an optimal external coupling of $\kappa_e/2\pi \approx 360\,$MHz (see Sec.\,\ref{sec:entangling-gate}).  We note that the above parameters and those that will follow are set according to the projected values of our system.

To load an atom to the cavity, atoms are first cooled in a magneto-optical trap (MOT), and then loaded and further cooled into an optical dipole trap using 1D polarization gradient cooling (PGC).
The trap is a far-detuned 1D optical lattice formed by a lower frequency TEM$_{00}$ mode, providing strong confinement in all three dimensions.

\subsection{Theoretical model}
\label{subsec: theoretical model}

\subsubsection{Qubit space}

The atomic qubit is encoded in two internal states of the ground-state manifold of the $^{87}$Rb atom. In our implementation, we choose $\lvert F = 1, m_F = \pm 1\rangle$ as the computational basis states (see Fig.~\ref{fig:UC-level-structure}(a)). Once this choice is made, all other atomic ground-state levels are regarded as leakage.
While, in principle, any two ground-state sublevels can serve as a qubit basis, our choice is motivated by the specific advantages and constraints of the cavity-QED platform. The selected atomic qubit states are energetically degenerate, therefore can couple selectively to the two orthogonal cavity modes. This allows us a straightforward cavity-based (and therefore $\sim10\,$ns-fast) implementation of the required set of quantum operations. This choice nevertheless involves tradeoffs. In typical neutral-atom architectures, the two $m_F=0$ hyperfine ``clock'' states are often preferred as the computational basis because of their resilience to magnetic noise and correspondingly longer coherence times. In our MBQC-based scheme, the atom performs only a short set of rapid entangling operations (see sec~\ref{sec:GS}), and is not required to support extremely long coherence times.  Accordingly, in this setting, the speedup enabled by having both atomic qubit states accessible to cavity-enhanced operations makes this tradeoff favorable.

The photonic qubit is encoded in a dual-rail basis, in which a single-photon excitation occupies one of two orthogonal optical modes, $|0_p\rangle , |1_p\rangle$ each with a well-defined spatial path, polarization, and temporal mode. In our implementation, these modes are realized primarily as two distinct spatial rails. Single-qubit operations are implemented by coherent mixing of the two rails using linear optical elements ~\cite{kok2007linear,rudolph2017optimistic,romero2024photonic}, while active optical components are used for fast routing.
A major advantage of our compound-qubit approach is that all entangling operations are based on direct photon--atom interactions, and do not rely on interference between independently generated photons such as fusion operations~\cite{bartolucci2023fusion}, therefore not requiring indistinguishable photons.

\subsubsection{Description of the model}
\label{subsubsec: model-description}
As described in Sec.~\ref{subsec: Physical layer - Unit-Cell}, the fundamental building block of our system is a 
$^{87}\mathrm{Rb}$ atom coupled to a Fabry--Pérot cavity. The atomic level structure relevant to the protocols considered here, corresponding to the $D_1$ line, is shown in Fig.~\ref{fig:UC-level-structure}(b). We denote the $5S_{1/2}$ ground-state manifold by $\mathcal{M}_g$, and the $5P_{1/2}$ excited-state manifold by $\mathcal{M}_e$. To characterize the optical coupling between these manifolds, we introduce the $q$-polarized atomic lowering operator:
\begin{equation}
\hat{D}_q = \sum_{\alpha\in \mathcal{M}_g}\sum_{\beta\in \mathcal{M}_e} c^{(q)}_{\alpha\beta} \ket{\alpha} \bra{\beta},
\end{equation}
where $q=0,\pm1$ correspond to $\pi$ and $\sigma_\pm$ polarized transitions, respectively. The coefficients $c^{(q)}_{\alpha\beta}$ are the dipole matrix elements, normalized to the value of $d$ --- the reduced dipole moment of the $D_1$ transition.

We model the atom--cavity unit cell as an open quantum system subject to dissipative processes, including photon leakage from the cavity and atomic spontaneous emission. The dynamics are described by the Lindblad master equation
\begin{equation}
\dot{\rho} = -i[\mathcal{H},\rho] + \mathcal{L}_{\kappa_i}[\rho]
+ \mathcal{L}_\gamma[\rho].
\end{equation}
Here the cavity dissipation is captured by the $\mathcal{L}_{\kappa_i}$ term with intrinsic rate $\kappa_i$, while atomic spontaneous emission is described by $\mathcal{L}_\gamma$ with field decay rate $\gamma/2\pi \approx 2.87\,\mathrm{MHz}$. Throughout this work, we set $\hbar=1$.

The coherent atom--cavity dynamics are generated by the Hamiltonian
\begin{equation}
\mathcal{H} = \mathcal{H}_a + \mathcal{H}_c + \mathcal{H}_{\mathrm{int}} + \mathcal{H}_{\mathrm{drive}}(t).
\end{equation}
The cavity Hamiltonian $\mathcal{H}_c$ includes  two degenerate polarization modes with the frequency $\omega_c$:
\begin{equation}
\mathcal{H}_c = \omega_c \left(a^\dagger_+ a_+^{\vphantom{\dag}} + a^\dagger_- a_-^{\vphantom{\dag}}\right),     
\end{equation}
where $a_+$ ($a_-$) is the annihilation operator for the $\sigma_+$ ($\sigma_-$) mode. The atomic Hamiltonian $\mathcal{H}_a$ describes the internal energy-level structure of the atom, as shown in Fig.~\ref{fig:UC-level-structure}(b). The atom-cavity interaction has the form of a generalized Jaynes--Cummings Hamiltonian, written in the rotating-wave-approximation:
\begin{equation}
\mathcal{H}_\text{int}=g \left(a_+^\dagger \hat{D}_+^{\vphantom{\dag}} + a_-^\dagger \hat{D}_-^{\vphantom{\dag}} + \text{h.c.}\right).
\end{equation}
Here, $g = E_0d$ is the atom--cavity interaction strength, with $E_0$ the cavity field amplitude at the position of the atom (see Eq.~\ref{eq: cavity field formula}). 
Using the geometrical and physical parameters discussed above, by placing the atom where $E_0$ is maximized, we obtain $g/2\pi \sim 360\,\mathrm{MHz}$.
We choose the atomic quantization axis to be aligned with the cavity axis, such that the cavity modes couple directly to the atomic $\sigma_\pm$ transitions.

We have performed a comprehensive analysis which, in addition to the atom--cavity dynamics described above, accounts also for the quantized atomic motion within the trapping potential. This entails both the kinetic contribution associated with the trapped atom and the spatial dependence of the atom--cavity coupling strength, $g(\mathbf{\hat{r}})$, arising from the inhomogeneous cavity field. 
While a detailed description of these effects is beyond the scope of the present work, our analysis indicates that they do not qualitatively alter the main results. 
Note that temporal and mode-dependent recoil effects give rise to infidelity in systems involving photonic Bell-state measurements~\cite{kikura2025taming,yu2026entanglement, apolin2026recoil}, and are not relevant in our setting.

Finally, the various protocols implemented in our system involve a classical coherent drive applied to selected atomic transitions, described by $\mathcal{H}_{\mathrm{drive}}(t)$. We leave the explicit form of this term unspecified, as it depends on the particular protocol under consideration.

It is useful to define a dimensionless quantity $C = g^2/\kappa \gamma$,  known as the cooperativity, which quantifies the strength of coherent atom--cavity interactions relative to the intrinsic loss mechanisms and therefore serves as a key figure of merit for the system.
Throughout this work, we use the notation $C$ to denote the specific cooperativity of the transition relevant to each protocol, avoiding the need to specify explicitly the various dipole matrix elements.

The interaction of single-photon pulses with the atom--cavity system is treated using the input--output formalism. We assume that the cavity couples to a continuum of external electromagnetic modes through its partially transmitting mirror. We introduce the input and output field operators in vector form as $\mathbf{a}_\text{in}(t)$ and $\mathbf{a}_\text{out}(t)$ respectively, and similarly for the cavity modes $\mathbf{a}(t)$. Each vector has two components corresponding to the two degenerate photonic modes of the system. These obey the relation~\cite{gardiner1985input,PhysRevLett.70.2273}:
\begin{equation}
\label{eq: input-output equation}
    \mathbf{a}_{\mathrm{out}}(t) = \mathbf{a}_{\mathrm{in}}(t) + \sqrt{2\kappa_e} \mathbf{a}(t).
\end{equation}
Owing to the symmetry of the cavity the choice of basis is not unique, and any linear combination of the two modes can be a valid and useful choice which satisfies the above equation. The polarization channels associated with these modes can be separated and routed into independent fibers or on‑chip waveguides using polarization‑selective optical components integrated into the surrounding photonic interface.

\subsection{Entangling gate}
\label{sec:entangling-gate}

The photon--atom CZ gate constitutes a central element of our architecture and serves as the primary building block for the generation of cluster states in MBQC. Our implementation employs a dual-rail encoding, in which each rail represents a distinct basis state of the photonic qubit. Although the CZ gate is symmetric with respect to the choice of control qubit, the atom constitutes the natural control in our implementation. The ability to perform arbitrary single-qubit rotations on the photonic mode enables straightforward conversion of the CZ gate into a CNOT gate by applying Hadamard operations to the photonic qubit.

\begin{figure}
    \centering
    \includegraphics[width=1.0\linewidth]{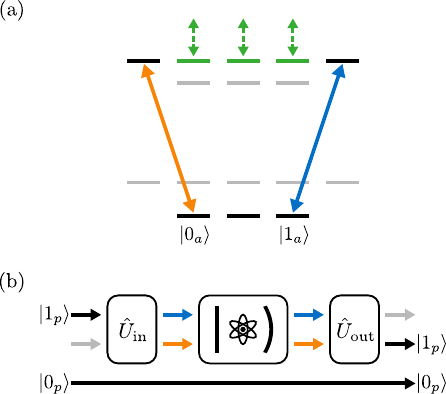}
    \caption{CZ gate description. (a) The atomic level structure shows the two transitions relevant to the gate operation. Each transition couples a qubit state to its corresponding excited state via $\sigma_-$ (orange) or $\sigma_+$ (blue), while transitions through the central excited state are suppressed by the level-engineering field, shown in green. (b) Schematics of the quantum channels involved in the interaction. The $\ket{0_p}$ rail does not interact with the atom, whereas the $\ket{1_p}$ rail interacts with a superposition of the two cavity modes defined by the unitary transformation $\hat{U}_\text{in}$. At the output the modes are transformed by $\hat{U}_\text{out}$. For the CZ gate, $\hat{U}_\text{in}$ is chosen such that the output channel is orthogonal to the input channel. }
    \label{fig: CZ-gates}
\end{figure}

The photon--atom CZ gate is implemented using a dual-rail variant of the Duan--Kimble scheme \cite{duan_kimble_cz}. The $|1_p\rangle$ rail is routed through the cavity and can interact with the atom, whereas the $|0_p\rangle$ rail bypasses the cavity and remains completely decoupled from the atomic state (see Fig.~\ref{fig: CZ-gates}). The key idea is that the phase acquired by a photon reflected from the cavity depends on the atomic qubit state, due to selective coupling of the $\sigma_\pm$ cavity modes, each to only one of the two atomic qubit states. In the atom--cavity configuration considered here, this condition is not naturally satisfied, because both qubit states are coupled not only to the edge excited states but also to the common excited state $|F'=2,m_F=0\rangle$ (see Fig.~\ref{fig:UC-level-structure}(b)). However, these unwanted transitions can be efficiently suppressed through the use of a properly designed control field~\cite{PRXQuantum.6.010340}. This level engineering scheme will be detailed in Sec.~\ref{subsec:le}. Here, we restrict our analysis to the four atomic states interacting with the cavity and are relevant to the operation of the gate.

When an incoming photon excites a cavity mode that is not coupled to an atomic transition, the reflection amplitude, which satisfies Eq.~(\ref{eq: input-output equation}), is
\begin{equation}
\label{eq:rc}
r_c= 1- \frac{2\kappa_e}{\kappa} ,
\end{equation}
which corresponds to the bare-cavity response. In the ideal limit $\kappa_e \rightarrow \kappa$, this expression approaches $r_c \rightarrow -1$. If, however, the photon excites a cavity mode that interacts with the atom \cite{CQEDreview}, the photon's reflection amplitude is given by
\begin{equation}\label{eq:ra}
r_{a}= 1- \frac{2\kappa_e}{\kappa}\left(\frac{1}{1+C}\right),
\end{equation}
which approaches $r_{a}\rightarrow 1$ in the limit of infinite cooperativity.

Choosing the $|1_p\rangle$ rail photon to couple to the $\sigma_-$ cavity mode, the photon--atom basis states are mapped such that $|1_p, 0_a\rangle \rightarrow r_a|1_p,0_a\rangle $ and $|1_p,1_a\rangle \rightarrow r_c|1_p,1_a\rangle $. The states corresponding to the non-interacting rail, $|0_p,0_a\rangle$ and $|0_p,1_a\rangle$, remain unchanged. In the ideal limit, this realizes the desired photon--atom CZ gate.

The difference between the two reflection amplitudes, i.e. asymmetric photonic loss, can be a major source of infidelity. For instance, the state $|1_p,+_a \rangle$ will be mapped to 
\begin{equation}
|1_p,+_a \rangle \ \rightarrow \ |1_p,-_a \rangle + \frac{r_c+r_{a}}{2}|1_p,+_a\rangle\: ,
\end{equation} 
where the second term is a coherent phase error. One way to reduce this error is to choose a cavity coupling $\kappa_e$ in which the scattering amplitudes are balanced, $r_c=-r_a$. This optimal coupling is given by
\begin{equation}
\kappa_e^{\text{opt}}=\kappa_i \sqrt{1+C_i},
\end{equation}
where $C_i = g^2/\kappa_i \gamma$ is the intrinsic cooperativity. In this case, the CZ gate efficiency can be conveniently expressed in terms of $C_i$. For an incoming $|+_p \rangle = (|0_p \rangle + |1_p \rangle)/\sqrt{2}$ photon state, and assuming the non-interacting $|0_p \rangle$ rail is ideal, the efficiency is given by:

\begin{equation}
    \eta=\frac{1}{2} \left(1 + 
    \frac{C_i^2}{\left(1 + \sqrt{1 + C_i}\right)^4}\right).
\end{equation}
For our system parameters, this yields an efficiency of $\eta \approx 98.4\%$. While optimizing the cavity output coupling already suppresses the error due to asymmetric photonic loss (with respect to the atomic qubit state), a more robust approach is to use the symmetrized version of the protocol, described below.

An inherent advantage of our configuration is that the degeneracy of the cavity modes enables several equivalent implementations of the CZ gate within the same system, depending on which cavity mode is excited. This mode can be chosen via optical components operating on the cavity input, represented by $U_\mathrm{in}$ in Fig.~\ref{fig: CZ-gates}(b). After the interaction, the two modes are transformed by $\hat{U}_\text{out}$, which, in the case of a single-sided cavity, corresponds to the backward-propagating transformation of $\hat{U}_\text{in}$. Choosing the photonic $|1_p \rangle$ rail to excite the  $\sigma_-$ cavity mode instead of the $\sigma_+$ mode, effectively exchanges the roles of the atomic $|0_p \rangle$ and $|1_p \rangle$ states.

A particularly appealing option is a symmetric CZ configuration \cite{PhysRevA.109.032602}, in which the incoming photon excites the linear $(\sigma_+ + \sigma_-)$ polarization, such that both atomic qubit states interact with the cavity mode. Following the interaction, the cavity photon will be in the 
orthogonal linearly polarized $(\sigma_+ - \sigma_-)$ mode, and reflected to a different output channel.
As before, the sign of the reflection amplitude is controlled by the atomic state. Here, the absolute values are the same for both states, such that $|1_p,0_a\rangle \rightarrow r_\text{sym}|1_p,0_a\rangle $ and $|1_p,1_a\rangle \rightarrow - r_\text{sym}|1_p,1_a\rangle $, where $r_\text{sym}$ is given by
\begin{equation}
r_\text{sym} = \frac{r_a-r_c}{2} = \frac{\kappa_e}{\kappa} \left(\frac{C}{1+C}\right).
\end{equation}
In this configuration, both atomic qubit states interact with the cavity symmetrically, which suppresses errors due to loss imbalance, regardless of the cavity and atom parameters.

Lastly, the $\ket{0_p}$ rail, which was assumed ideal so far, also suffers loss due to optical propagation and routing elements. Any imbalance between the total losses of the two rails leads again to an effective phase error. To mitigate this effect, the photonic loss must be balanced according to the full optical path, taking into account all entangling gates the photonic qubit undergoes. One way to achieve this is to place some of the entangling operations on the $ |0_p\rangle$ photonic rail.
Logically, this operation is equivalent to conjugating these CZ gates by photonic bit flips, i.e. $X_p\,\mathrm{CZ}\,X_p$.

\subsection{State preparation and measurement (SPAM)}
\label{subsec: SPAM}
Preparation of the atom in the qubit subspace can be achieved via a measurement-based state preparation: a photon--atom Bell state \cite{PhysRevLett.96.030404,wilk2007single} is first generated using vSTIRAP, and the photonic qubit is then measured in the desired basis, thereby projecting the atom onto the corresponding target state.

\begin{figure}[h]
\centering
\includegraphics[width=\linewidth]{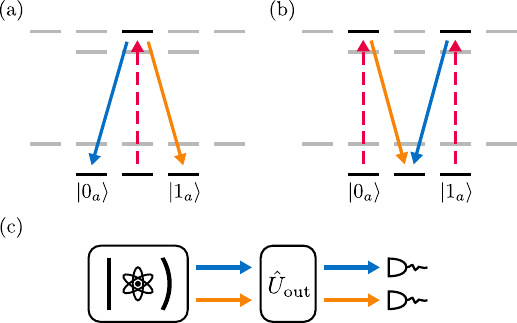}
\caption{Atomic state preparation and measurement protocols. In (a) the excitation of the driven atom (red arrows) from $|F=1,m_F=0\rangle$ is followed by a coherent transition (blue/orange) into the atom qubit--space. Similarly in (b), the driven atom transitions out of the qubit-space. Both processes are schematically summarized in (c), which shows the two emitted cavity modes (blue/orange). A unitary operation $\hat{U}$ dictates the measurement basis.}
\label{fig:vprep_d1}
\end{figure}

The protocol begins by optically pumping the atom to the $\lvert F=1,m_F=0\rangle$ state. A $\pi$-polarized pulse, resonant with the $D_1$ transition $F=1  \rightarrow F'=2$ then drives the atom to the excited state $\lvert F'=2,m_F=0\rangle$, as can be seen in Fig.~\ref{fig:vprep_d1}(a). The excited state is symmetrically coupled via the cavity to the two qubit states $\lvert F=1,m_F=\pm1\rangle$, resulting in the emission of a cavity photon.
The photon's polarization is thereby entangled with the atomic qubit, yielding the Bell state:
\begin{equation}
\lvert \psi_p,\psi_a \rangle \ = \  \frac{1}{\sqrt{2}} \left( \lvert  \sigma_+ ,0\rangle + \lvert \sigma_- ,1\rangle\right) ,
\end{equation}
where the $\lvert \sigma_\pm \rangle$ photon states correspond to the outgoing fields of the $a_\pm$ cavity modes.
Measurement of the outgoing photon projects the atom onto the corresponding qubit state. In particular, measuring the photon in the $X$ basis prepares the atom in either $\lvert +\rangle$ or $\lvert -\rangle$. For cluster-state generation, both outcomes can be accepted, since they differ only by a known Pauli-frame update and therefore do not require post-selection. The preparation is naturally heralded: successful detection of the emitted photon signals successful state preparation, while failed attempts can be identified and discarded. More generally, by choosing the photonic measurement basis appropriately, one can prepare arbitrary atomic-qubit states, such as $T$-state, without requiring direct single-qubit rotations on the atom. The choice of measurement basis can be achieved using optical components, represented by $\hat{U}$ in Fig.~\ref{fig:vprep_d1}(c), that operate on the cavity output.

Atomic state measurement is performed by coherently mapping the atomic qubit onto a photonic qubit encoded in the cavity output modes. In this way, atomic-state measurement is reduced to photonic measurement. Similar to state preparation, atomic-state measurement is also based on a vSTIRAP process. The atom starts in the computational subspace and is then driven by a $\pi$-polarized pulse to the excited states $\ket{F'=2,m_F=\pm1}$. These excited states are coherently coupled by the cavity modes to the state $\ket{F=1,m_F=0}$ (see Fig.~\ref{fig:vprep_d1}(b)). A successful operation emits a single photon while transferring the atom to $\ket{F=1,m_F=0}$. The polarization of the emitted photon carries the atomic-qubit information and can subsequently be measured in any desired basis. This photon may also be routed into subsequent gate operations before being detected. In the following sections, we will refer to this process as STAP (State Transfer from Atom to Photon).

The atomic state preparation and measurement protocols are complementary by design, which is expected to improve overall performance. First, both protocols use the same optical path to the detectors, eliminating relative phase ambiguities between preparation and measurement photons that would otherwise lead to coherent errors. Second, the final atomic state following a successful measurement coincides with the required initial state for atomic preparation. This feature is naturally compatible with MBQC, where qubits undergo a repeated sequence of preparation, entanglement, and measurement.

At the unit-cell level, the efficiency of the preparation and measurement protocols is determined by the probability that the atom-cavity Raman process emits a photon to the desired output mode. For atomic measurement, the emission efficiency is given by \cite{CQEDreview}
\begin{equation}
\label{eq: effiency of vprep}
\eta \simeq \frac{\kappa_e}{\kappa}\frac{C}{C+1}.
\end{equation}
The ratio $\kappa_{e}/\kappa$ gives the probability that a cavity photon exits through the designated output mode, while $C/(1+C)$ gives the probability of atomic emission into the cavity mode rather than into free space. For the state-preparation protocol, the corresponding expression is obtained by formally replacing $C \rightarrow 2C$, due to the two cavity-enhanced decay paths. The end-to-end efficiencies of the full protocols additionally include propagation loss, routing loss, and detector inefficiency. Using the projected cavity-QED parameters presented in Secs.~\ref{subsec: Physical layer - Unit-Cell} and~\ref{subsec: theoretical model}, the cavity-limited contribution corresponds to efficiencies of $\eta\approx 96.9\%$ for state preparation and $\eta\approx 97.6\%$ for atomic state measurement (taking into account the different $C$ of the two processes). 

A major source of error in the SPAM protocols is multi-photon emission from the cavity. Ideally, each successful attempt emits exactly one photon and leaves the atom in the desired final state. However, in this configuration the atom may be re-excited within the same optical pulse after a photon has already been emitted, leading to multi-photon events and/or residual population outside the intended state. Such multi-photon events can produce unheralded preparation or measurement errors if only one of the emitted photons is detected. These errors can be significantly mitigated using multi-stage STIRAP schemes and optimized pulse shapes that suppress re-excitation pathways while preserving high efficiency, as confirmed by our analysis, which is beyond the scope of this manuscript.

\subsection{Single-photon generation}

The atom--cavity unit cell can also serve as a single-photon source based on vSTIRAP, in a $\Lambda$ configuration \cite{kuhn2010cavity}. The atom is initially prepared in the state $\ket{F=2,m_F=2}$ by optical pumping. A $\pi$-polarized pulse drives the transition to the excited state $\ket{F'=2,m_F=2}$, which is then coherently transferred to $\ket{F=1,m_F=1}$ while emitting a photon into the $\sigma_+$ cavity mode. Since the initial and final states differ in energy by approximately $6.8\,\mathrm{GHz}$ (see Fig.~\ref{fig:UC-level-structure}(b)), re-excitation is strongly suppressed, enabling the generation of high-purity single-photon pulses. The efficiency, which is given by the same expression as for state preparation in Eq.~(\ref{eq: effiency of vprep}), evaluates to $\eta \approx 98.4\%$. To calculate this we used the cooperativity relevant to the transition in the process above.

The temporal mode of the emitted photon is controlled by the shape of the drive pulse. However, the cavity-QED parameters set a lower limit on the pulse length. If the drive is too fast, the atom--cavity dynamics can no longer follow adiabatically, the temporal mode of the emitted photon is distorted, and the generation efficiency is reduced. The relevant requirement is that the pulse width $\tau$ remains long compared with the slowest response time of the system,
\begin{equation}
\tau \gg \max\!\left(\frac{1}{\kappa}, \frac{\kappa}{g^2}\right),
\end{equation}
which is $1/\kappa$ in the strong-coupling regime, and $\kappa/g^2$ in the fast cavity regime. 
Note that as long as this limit is met, neither the exact shape of the pulse nor its temporal purity affect the entangling gate fidelity~\cite{PRXQuantum.6.010340}. Using the parameters presented in Secs.~\ref{subsec: Physical layer - Unit-Cell} and~\ref{subsec: theoretical model}, this requirement is satisfied for $\tau \sim 10\,\mathrm{ns}$. A similar photonic pulse-length requirement applies across the platform, including in atom state preparation, measurement, and the CZ gate. In practice, this condition is naturally satisfied, since the same unit cell is used for all protocols, and therefore operates on a common set of characteristic timescales. Furthermore, the fact that all modules are identical enables changing their roles dynamically, depending on higher-level design considerations.

\subsection{Level engineering}\label{subsec:le}
Here we use the term \emph{level engineering} to denote the controlled reshaping of the atom excited-state structure by an auxiliary control field. In our implementation, this is achieved optically, by coupling the $5P_{1/2}$ manifold to the higher-lying $6S_{1/2}$ manifold using a $1324\,\mathrm{nm}$ laser~\cite{PRXQuantum.6.010340}. This manifold has a similar hyperfine structure to the $5S_{1/2}$ (see Fig.~\ref{fig:UC-level-structure}(b)), and a linewidth $\tilde{\gamma}/2\pi \approx 1.73\text{MHz}$. Specifically, we apply a $\pi$-polarized field resonant with the $F'=2 \rightarrow F''=1$ transition. This hybridizes the $m_F=-1,0,1$ states of the two manifolds, producing a dressed-state splitting of order $\pm \Omega$, where $\Omega$ is the Rabi frequency induced by the control field. By contrast, the $m_F=\pm 2$ states are only weakly affected, acquiring a small shift of order $-\Omega^2/\delta$ due to their off-resonant coupling to the $F''=2$ levels, where $\delta \approx 1.6\,\mathrm{GHz}$ is the hyperfine splitting of $6S_{1/2}$.

For the CZ gate, the goal of this procedure is to suppress the unwanted transitions from the qubit subspace to the $|F'=2,m_F=0\rangle$ level, while retaining the desired coupling to $|F'=2,m_F=\pm2\rangle$. In the presence of the level-engineering field, the corresponding cooperativity of the unwanted transition is effectively reduced to $C \rightarrow C\,\gamma/\gamma_{\mathrm{eff}}$ with ${\gamma_{\mathrm{eff}}=\gamma+\Omega^2/\tilde{\gamma}}$, similar to what occurs in electromagnetically induced transparency configuration~\cite{Finkelstein_2023}. Consequently, the unwanted atomic transition becomes effectively decoupled once $C\gamma/\gamma_{\mathrm{eff}} \ll 1$, or equivalently when
\begin{equation}
\Omega \gg \sqrt{C\gamma\tilde{\gamma}}.
\end{equation}
With our working parameters, $\Omega/2\pi \sim 200\,\text{MHz}$ will satisfy this condition.
Such a level-engineering scheme has already been demonstrated experimentally in our system, and will be reported in a forthcoming publication.

\section{Fault tolerant memory channel}

With the physical properties of individual qubits and gates established, we now turn to how these components are assembled into a functional quantum computing architecture. In this section we describe the hardware elements, cluster-state generation schemes, and fault-tolerance framework that together connect the physical layer to reliable logical computation.

\subsection{Hardware architecture elements}

\begin{figure*}
\centering
\includegraphics[width=\linewidth]{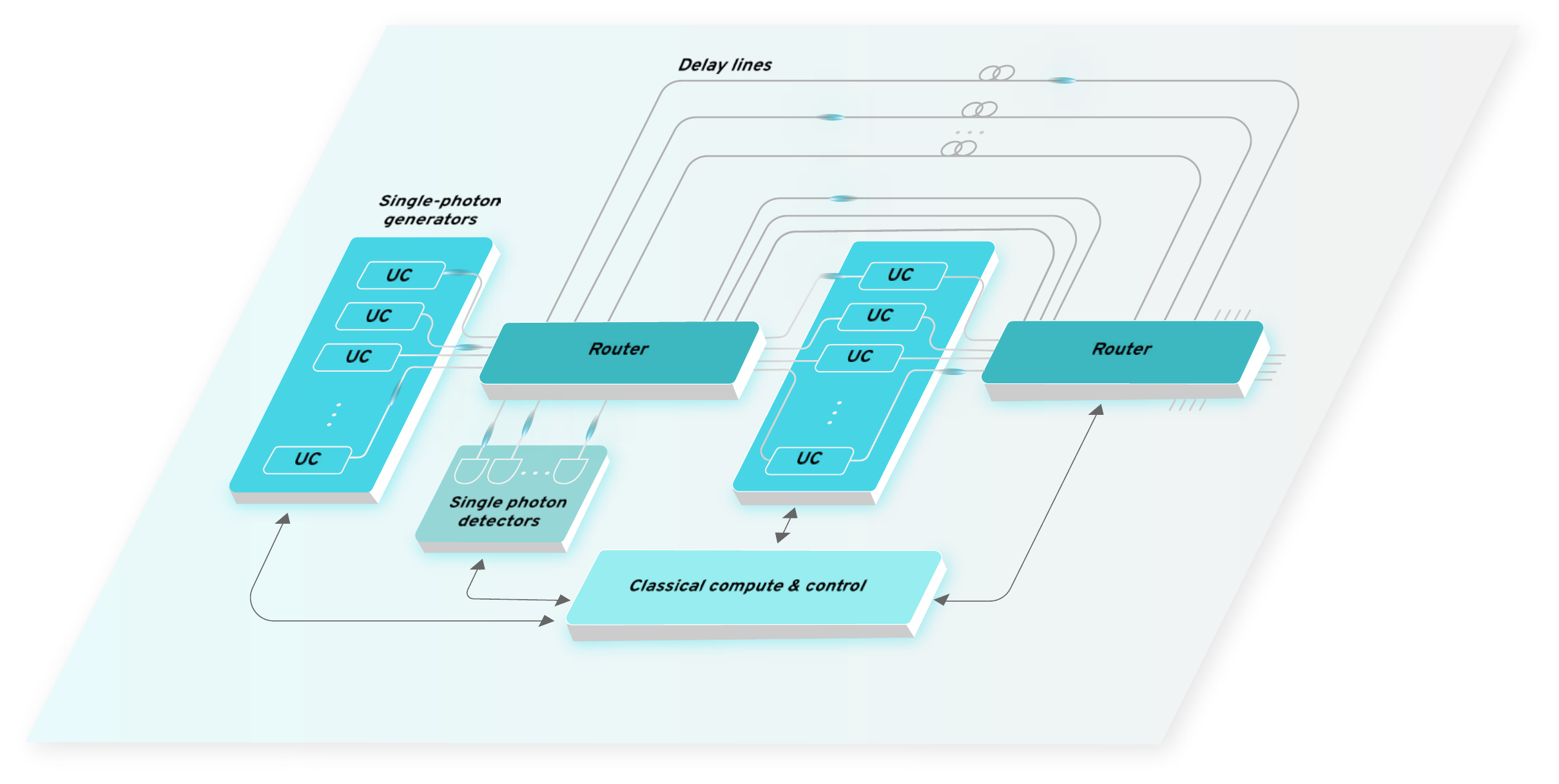}
\caption{ High-level hardware architecture concept of a photon--atom MBQC system. Unit cells generate photonic qubits, mediate photon--atom entanglement, and perform atomic SPAM operations. Reconfigurable routing manifolds and delay lines establish the temporal and spatial connectivity required for cluster-state generation. Photons are ultimately measured by single-photon detectors, while a classical control and processing layer coordinates system operation and performs real-time error-correction decoding based on detector outcomes.
}
\label{fig:HW_concept}
\end{figure*}

Fig.~\ref{fig:HW_concept} illustrates the principal architectural elements required to implement measurement-based quantum computation (MBQC) on compound photon--atom quantum hardware.

As described in Sec.~\ref{subsec: Physical layer - Unit-Cell}, unit cells (UCs) implement entangling gates, generate single photons, and perform state preparation and measurement (SPAM) on trapped atoms. Each UC has one input and one output, and UCs are interconnected by routing manifolds that dynamically realize the connectivity patterns required for cluster-state generation. These cluster states can be generated in different ways depending on atomic-resource usage, error budget, computational throughput, and other architectural trade-offs. Sec.~\ref{sec:GS} presents two representative generation schemes.

Although the bipartite photon--atom cluster states considered here are symmetric with respect to the two qubit types, their generation is most naturally described from the photonic perspective. A typical {\it photon cycle} consists of: (1) single-photon generation; (2) a sequence of photon--atom controlled-phase (CZ) gates; and (3) photon measurement in the X basis using single-photon detectors (SPDs).

After traversing a UC, a photon may be routed to another UC, a delay line, or an SPD. The routing configuration changes between photon cycles and, when delay lines are used, may also change within a cycle. Each atom participates in multiple photon cycles. Once all entangling operations involving a given atom have been completed, its state is transferred to a photon, as described in Sec.~\ref{subsec: SPAM}. The photon is then routed to an SPD for measurement.

The classical control system manages routing, photon generation, atom preparation and measurement, and orchestration tasks such as monitoring and diagnostics. In parallel, the SPDs produce a large stream of measurement data, which is processed into error-correction syndromes and supplied to real-time decoding algorithms running on high-performance classical hardware, as discussed below.

\subsection{Fault tolerance and MBQC}

Fault-tolerant quantum error-correcting codes allow quantum computers to compute reliably despite imperfections in qubits, gates, and measurements. In a fault-tolerant scheme, errors are repeatedly detected and corrected so that they neither propagate uncontrollably nor accumulate beyond the code capacity. When the physical error rate is below threshold, increasing the code size can in principle suppress logical errors to the level required for large-scale computation, thereby connecting the noisy physical layer to reliable logical computation.

The basic fault-tolerant primitive is a logical memory channel: a protected identity operation implemented either in the circuit-based (CBQC) or measurement-based (MBQC) model~\cite{terhal2015quantum}. In matter-based architectures, this primitive is usually realized in CBQC by repeated stabilizer measurements on persistent qubits. These measurements extract error syndromes while preserving the encoded logical information.

Photon-rich architectures are more naturally described in MBQC. Since photons propagate, cannot be measured repeatedly, and cannot be stored for long times without loss, computation is performed by preparing a large entangled cluster state and measuring it layer by layer~\cite{briegel2009measurement}. Each measured layer is consumed, while the encoded quantum information is teleported forward, and products of measurement outcomes provide the syndrome information needed for fault tolerance.

\begin{figure*}
    \centering
    \stackinset{l}{2pt}{t}{2pt}{(a)}{%
      \includegraphics[width=0.45\linewidth]{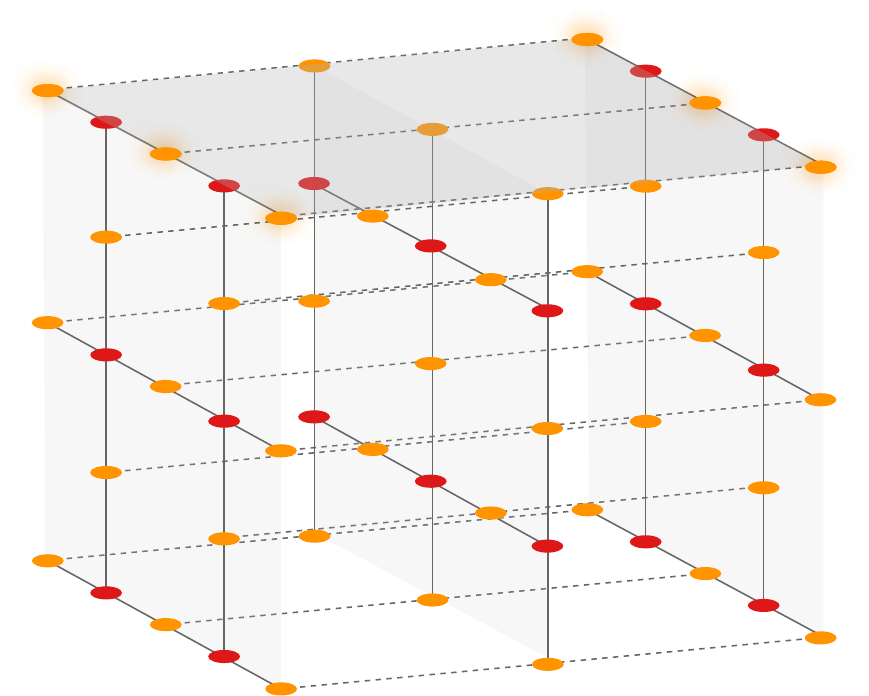}%
    }%
    \hfill
    \stackinset{l}{2pt}{t}{2pt}{(b)}{%
      \includegraphics[width=0.45\linewidth]{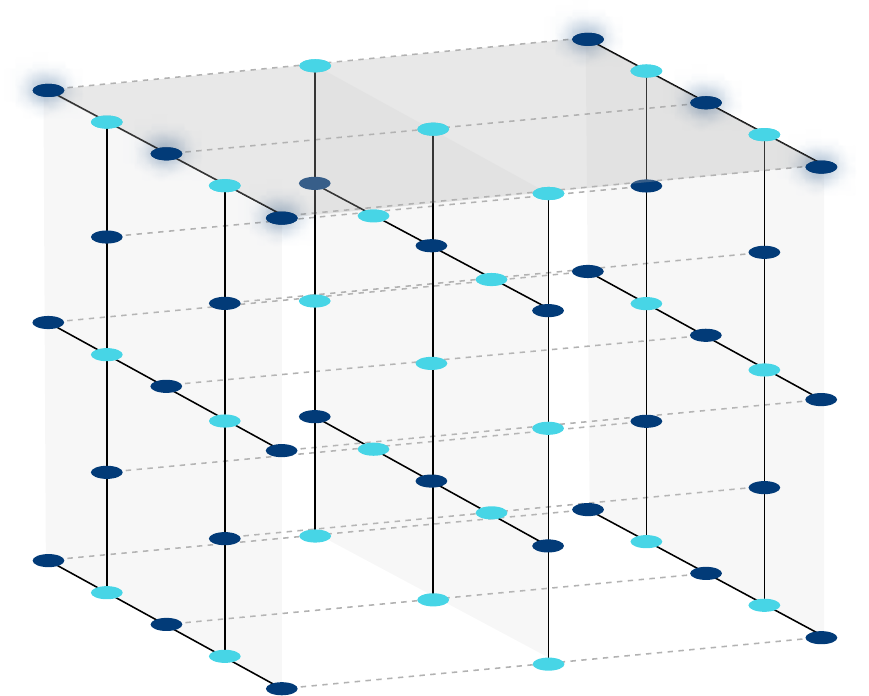}%
    }%
    \caption{A $d=3$, $t=3$ RHG construction starting and ending with dual layers, corresponding to logical $\ket{0}_{L}$ initialization and measurement. Circles denote qubits. Dashed gray lines indicate CZ gates in the time direction, while solid gray lines indicate CZ gates within each spatial layer. The number of data qubits along each spatial row or column is $d$, the error-correction code distance, and $t$ is the number of spatial layers stacked along the time direction. The horizontal shaded cap indicates a logical correlation surface, whose value is determined by the combined $X$-basis measurement outcomes of the glowing qubits. (a) Generic data--SL coloring: orange qubits are data qubits and red qubits are syndrome-like (SL) qubits. (b) Bipartite photon--atom coloring: cyan qubits are photons and dark-blue qubits are atoms. In the configuration shown, the correlation surface is supported on atoms, although the complementary photon-supported surface can be chosen instead. See Appendix~\ref{App:RHG-general} for details.}
\label{RHGconstruct}
\end{figure*}

Foliation provides the bridge between these two pictures~\cite{bolt2016foliated}: repeated syndrome-measurement rounds in CBQC are represented in MBQC as successive measurements of a three-dimensional cluster state, through which logical information is teleported forward while error information is accumulated and decoded. In this representation, \textit{data} qubits carry the propagated logical information, while \textit{syndrome-like} (SL) qubits are measured to produce the error-correction checks. The canonical example is the Raussendorf--Harrington--Goyal (RHG) lattice~\cite{raussendorf2007topological,fujii2015quantum}, which realizes the surface code as a three-dimensional cluster state, as shown in Fig.~\ref{RHGconstruct}. Its high iid measurement loss threshold of approximately $25\%$, low node degree, compatibility with optimized decoders, and well-established framework for logical operations make it a natural candidate for our architecture. Appendix~\ref{App:RHG-general} reviews the RHG construction, foliation structure, logical encoding, and error-correction procedure.

\subsection{Generation schemes for compound photon--atom hardware}\label{sec:GS}

A generation scheme specifies how the target cluster state, here the RHG lattice, is assembled from the native hardware operations. The physical qubits are atoms and photons, and the allowed operations are atom initialization, photon generation, photon--atom CZ gates, photon measurement, and STAP. The STAP operation transfers the state of an atom to an emitted photon, for example to enable atomic measurement (see Sec.~\ref{subsec: SPAM}). Photons may also be routed through switches and, when required, through single-qubit gates. Throughout this work, we refer to all of these elementary operations collectively as \textit{physical gates}. We model all gates as lossy except atom initialization, where loss is heralded and corrected before any interaction. Optical path loss between gates is assigned to the subsequent gate.

\begin{figure*}[t]
    \centering

    \begin{tabular}{@{}c@{\hspace{0.04\textwidth}}c@{\hspace{0.04\textwidth}}c@{}}
        \includegraphics[width=0.27\textwidth]{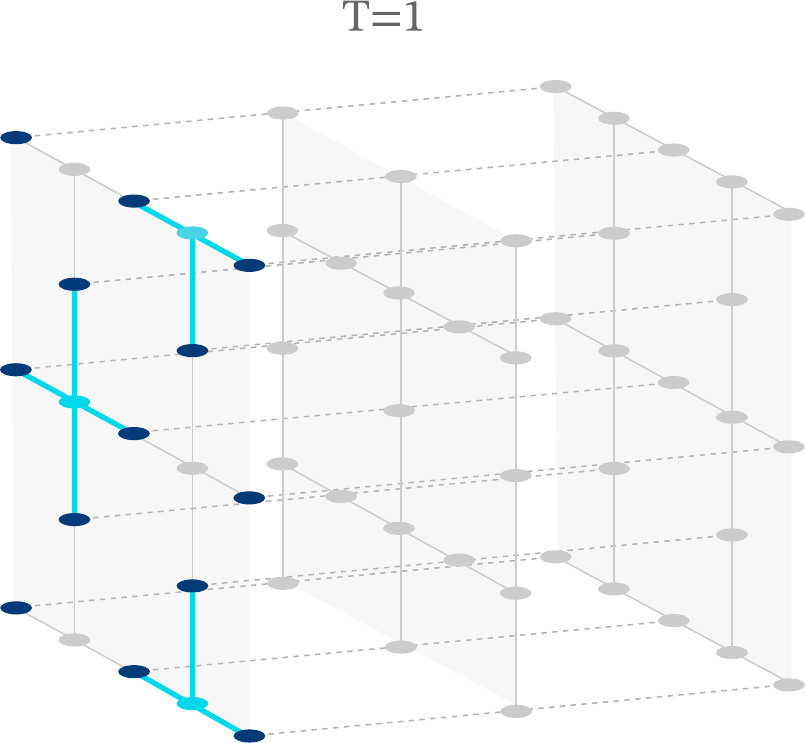} &
        \includegraphics[width=0.27\textwidth]{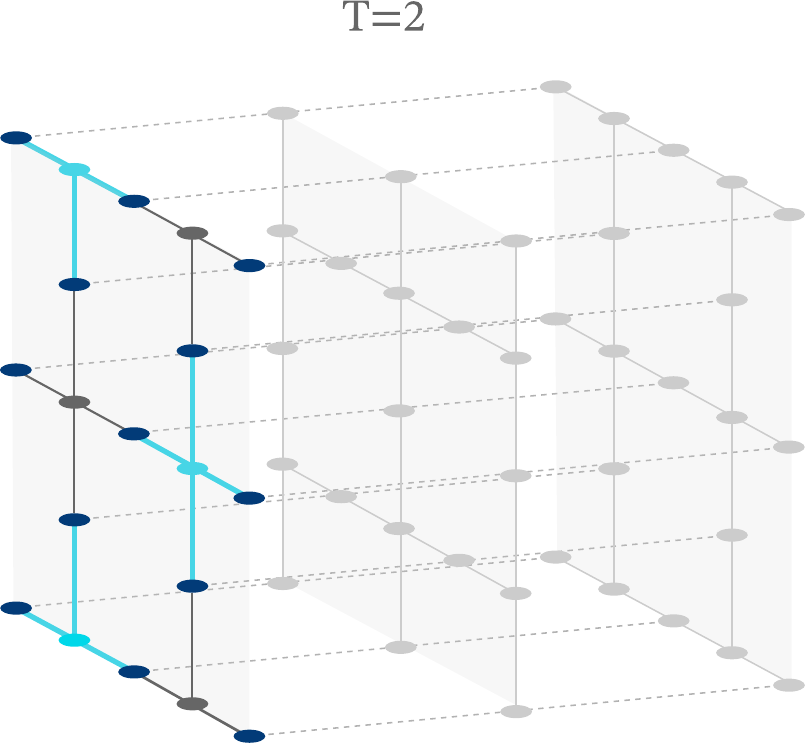} &
        \includegraphics[width=0.27\textwidth]{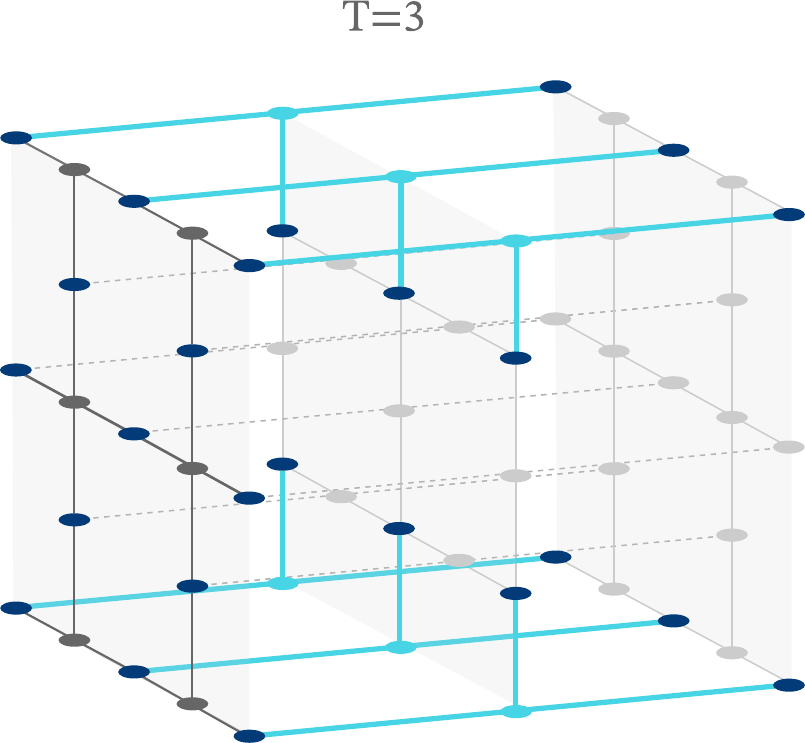}
        \\[0.8em]

        \includegraphics[width=0.27\textwidth]{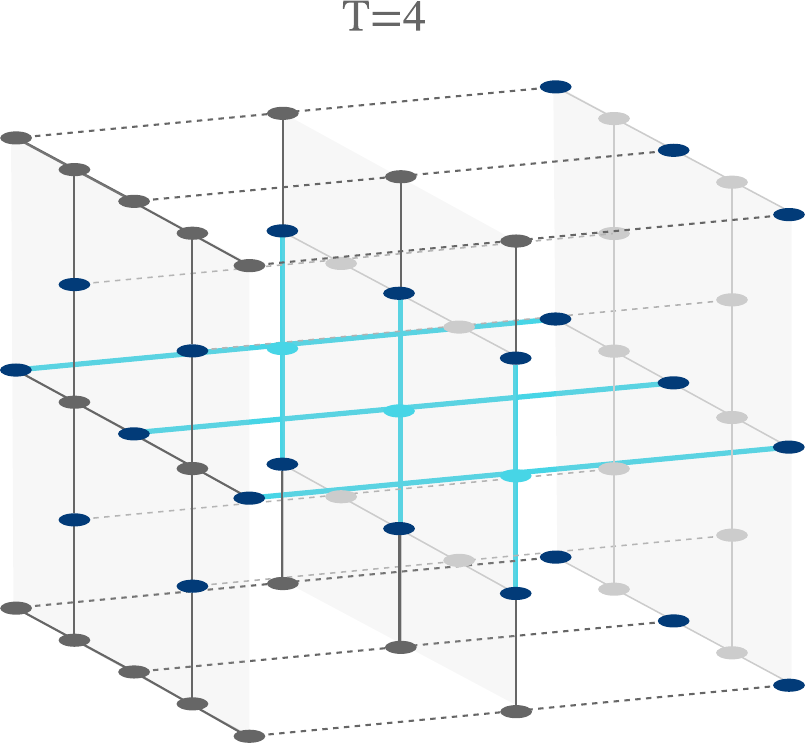} &
        \includegraphics[width=0.27\textwidth]{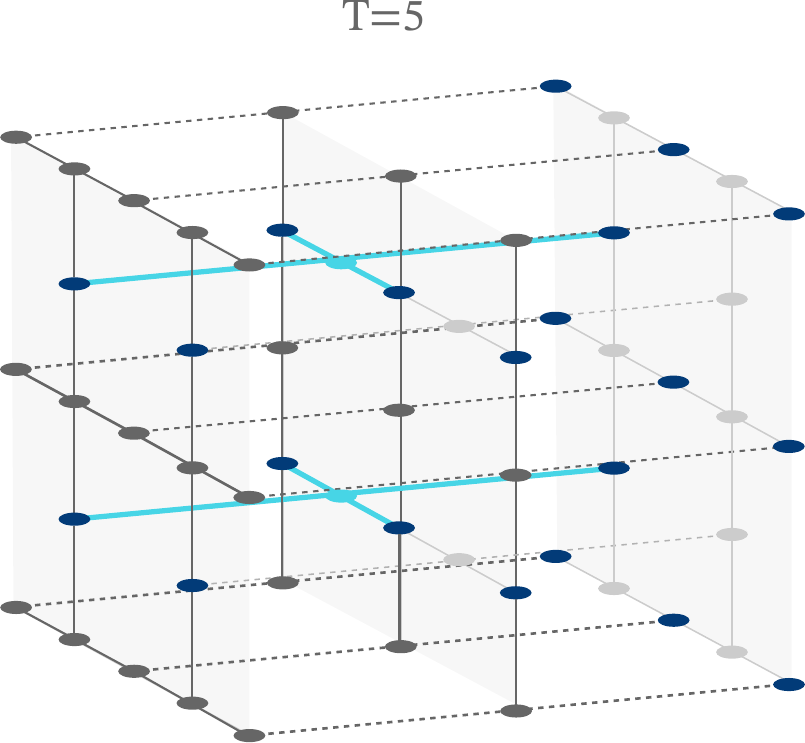} &
        \includegraphics[width=0.27\textwidth]{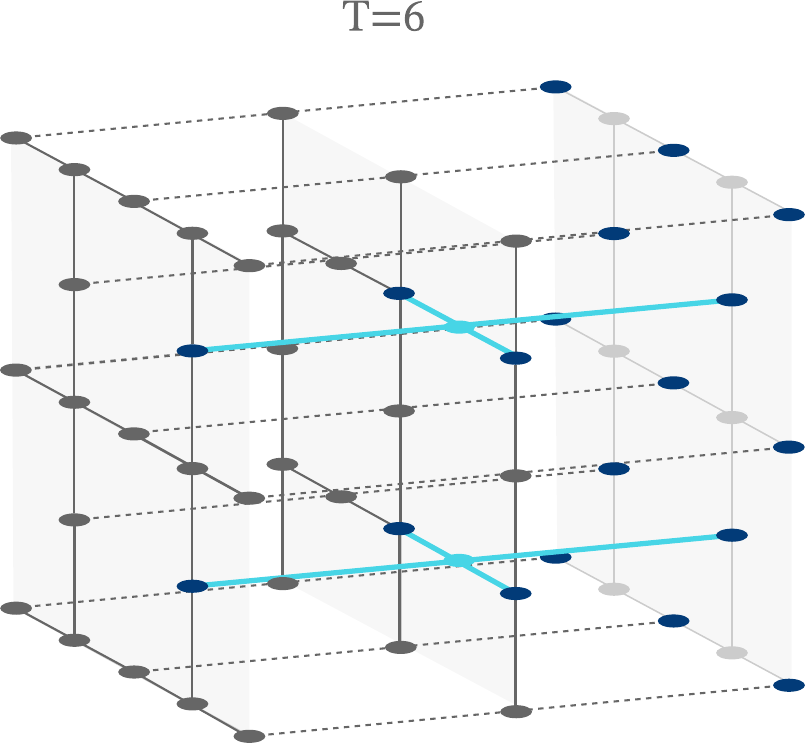}
    \end{tabular}

    \vspace{0.4em}

    $\xrightarrow{\hspace{0.8\textwidth}}$\\[-0.2em]
    \textit{Time}

    \caption{Time progression of the bipartite generation scheme. Time advances from left to right within each row and from the upper row to the lower row. Six consecutive time steps are shown. At each step, three adjacent RHG layers are displayed, labeled from left to right as $n$, $n+1$, and $n+2$. Empty lattice sites are shown in light gray, atoms in dark blue, and photon qubits in cyan. Cyan edges indicate CZ gates performed in the current time step, while black qubits and edges indicate measured qubits and previously performed CZ gates. Because each atom participates in at most one CZ gate per time step, the intra-layer CZ gates in layer $n$ are completed over two photon cycles, $T=1,2$. The data qubits in layer $n+1$ then perform four CZ gates over $T=3,\ldots,6$: two intra-layer gates within layer $n+1$ and two inter-layer gates connecting to layers $n$ and $n+2$. The SL qubits in layer $n+2$ are then introduced as photons, and the sequence repeats.}
    \label{fig: bipartite-scheme}
\end{figure*}

We use a \textit{bipartite generation scheme}, assigning one part of the cluster-state graph to photons and the other to atoms, as shown in Fig.~\ref{RHGconstruct}(b). Each bulk photon is initialized, interacts with four atoms via CZ gates, and is measured. Atomic qubits participate in the corresponding CZ interactions and are measured indirectly: their final states are first transferred to emitted photons using STAP and then measured. The gate ordering is constrained by the different operational constraints of photons and atoms. Photons complete their CZ interactions in a single pass, whereas atoms can participate in only one CZ gate at a time. Based on these gate-scheduling constraints, we implement the bipartite generation scheme approximately layer by layer. Two RHG layers require about six photon cycles, where one cycle is the interval in which a photon completes all of its gates, as shown in Fig.~\ref{fig: bipartite-scheme}. The scheme uses $O(d^{2})$ active atoms, with measured atoms recycled in later steps. The parameter $d$ denotes the number of qubits along each spatial row or column and sets the error-correction distance of the RHG lattice.

After STAP, the atom is left outside the qubit subspace in the initial state used for the atomic state-preparation protocol. To recycle an atom, it is then reinitialized in a fully heralded process to $\lvert+\rangle$ or $\lvert-\rangle$, as described in Sec.~\ref{subsec: SPAM}. The additional delay is not limiting, because unmeasured atoms can serve as short-term memory while recycled atoms complete reinitialization. This short-term memory also provides flexibility in the generation schedule: photon generation and routing can be delayed until the relevant atomic resources are ready, relaxing the required routing and switching rates.

\begin{figure*}[t]
    \centering

    \begin{tabular}{cccc}
        \includegraphics[width=0.18\textwidth]{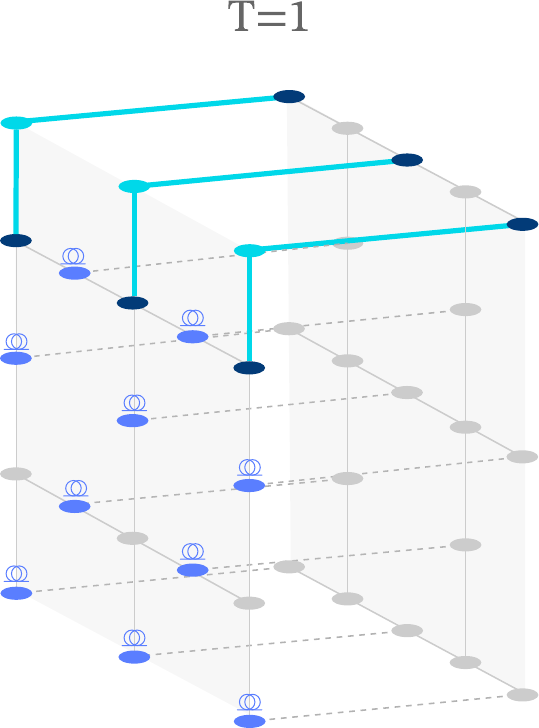} &
        \hspace{0.03\textwidth}
        \includegraphics[width=0.18\textwidth]{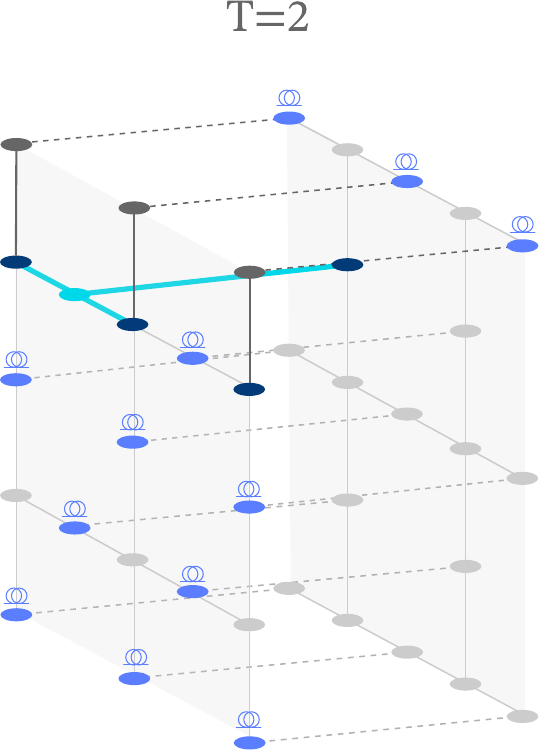} &
        \hspace{0.03\textwidth}
        \includegraphics[width=0.18\textwidth]{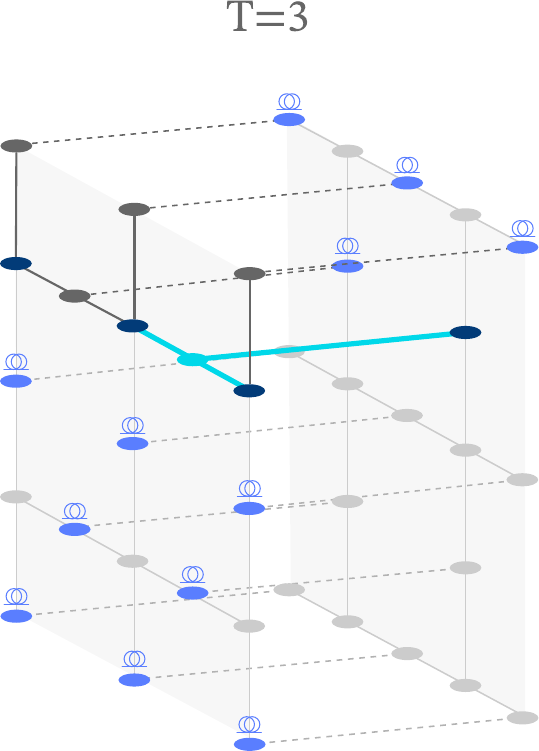} &
        \hspace{0.03\textwidth}
        \includegraphics[width=0.18\textwidth]{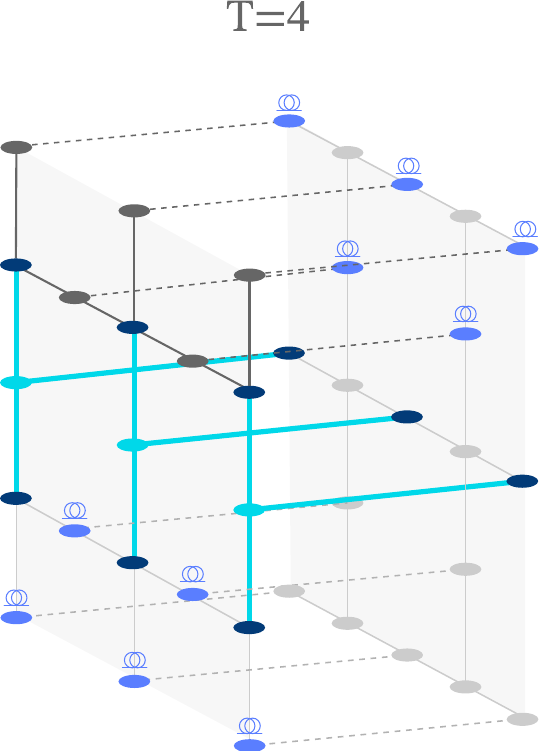}
    \end{tabular}

    \vspace{0.4em}

    $\xrightarrow{\hspace{0.8\textwidth}}$\\[-0.2em]
    \textit{Time}

    \caption{Time progression of the interleaved generation scheme. Time advances from left to right. Four consecutive time steps are shown, each involving two adjacent RHG layers: layer $n$ on the left and layer $n+1$ on the right. Empty lattice sites are shown in light gray, atoms in dark blue, and photon qubits in cyan. Cyan edges indicate CZ gates performed in the current time step, while black qubits and edges indicate measured qubits and previously performed CZ gates. Photons stored in delay lines are shown in purple, with delay-line symbols above them. Data photons from layer $n$ perform CZ gates with neighboring atomic SL qubits and atomic data qubits in layer $n+1$. The layer-$n+1$ atomic data qubits are then transferred to photons and routed into delay lines.}
    \label{fig:interleaving}
\end{figure*}

Inspired by interleaving ideas from Refs.~\cite{bombin2021interleaving, bartolucci2023fusion}, the atomic space overhead can be reduced from $O(d^2)$ to $O(d)$ using the  \textit{interleaved generation scheme} illustrated in Fig.~\ref{fig:interleaving}. This scheme combines STAP operations, which transfer atomic states to photons before all required CZ gates are complete, with delay lines that store the emitted photons between subsequent CZ interactions. The lattice is generated row by row, so only $O(d)$ atoms are active at any time. Conceptually, once an atom in row $r$ completes its role, it is reused in row $r+1$. Physically, this is implemented by reconfiguring the optical switches. To avoid delays due to atomic reinitialization, one may use $O(d)$ additional atoms, prepared in parallel during row generation.

As shown in Fig.~\ref{fig:interleaving}, each row of SL qubits (defined in Appendix~\ref{App:RHG-general}) is initialized as atoms before its first CZ interaction, remains atomic throughout its CZ gates, and is then measured, freeing the atoms for reuse. The data-qubit sequence is best described in terms of two adjacent layers. Data qubits from layer $n$ exit the delay line as photons and interact both with the neighboring SL row in layer $n$ and with the corresponding row of data qubits in layer $n+1$; the latter are initialized as atoms before these inter-layer CZ gates. The layer-$n$ data qubits are then measured as photons, while the atomic layer-$n+1$ data qubits are converted to photons and routed into delay lines. The freed atoms are reused in the next row. This sequence repeats row by row and layer by layer. The initial layer in Fig.~\ref{fig:interleaving} should be understood as beginning from the photonic stage of the cycle.

This reduction in atomic footprint comes at the cost of optical memory. The scheme requires one delay line per data qubit, each of length $O(d)$, since a layer contains $d$ rows and each row requires constant processing time. Thus, the $O(d)$ reduction in active atoms is exchanged for an $O(d)$ delay-line overhead. Since delay lines are lossy, this can increase photon loss and the logical error rate, especially at larger code distances.

\subsection{Error model} \label{sec: EM}

The compound architecture considered here carries quantum information jointly in photonic and atomic qubits, which are subject to different noise mechanisms and error-propagation pathways. The error model must therefore distinguish between photonic and atomic degrees of freedom while also accounting for correlated errors arising from photon--atom interactions.

Photonic qubits are modeled as lossy during propagation, cavity coupling, and measurement. We neglect photonic Pauli noise in this work because photon loss is expected to be the dominant error mechanism~\cite{psiquantum2025manufacturable,aghaee2025scaling,chan2025practical}. In addition, our photon--atom CZ protocol does not rely on multi-photon interference between indistinguishable photons, reducing sensitivity to the mode-mismatch errors that are central in linear-optical photonic schemes. Atomic loss and intrinsic atomic noise are also neglected, as they are expected to be much rarer than photon loss, and are left for future study. Nevertheless, photon loss can still induce effective atomic errors. Atomic readout is performed by transferring the atomic state onto a photonic one before measurement, so loss of the photon acts as a readout error. Similarly, photon loss during a photon--atom CZ interaction can remove an intended bond and induce correlated Pauli errors on neighboring atoms.

We distinguish between \textit{physical qubit loss}, such as photon loss in an optical path~\cite{barrett2010fault,fowler2013coping,ghosh2013understanding,ghosh2015leakage}, and \textit{bond loss}, in which an intended entangling operation fails~\cite{li2010fault,auger2018fault}. In our architecture, physical photon loss can induce bond loss: if a photon is lost before completing its scheduled CZ gates, the remaining bonds involving that photon are absent. These missing bonds distort neighboring stabilizers, randomize syndrome information, and corrupt the associated correlation surfaces, an effect we call \textit{bond-loss propagation}. Although known erasures can be partially handled by modifying the affected checks, bond-loss propagation requires an explicit loss-aware model because it can corrupt neighboring checks without producing a direct measurement signature. See Appendix~\ref{App:BondLoss} for a stabilizer-level explanation.

Many approaches approximate bond loss by introducing independent Pauli errors on qubits that participate in entangling gates with a lost qubit. Although simple, such approximations generally overestimate the damage caused by a missing bond~\cite{whiteside2014upper,whiteside2017classical, Perrin2025quantumerror}. More refined loss-aware methods condition the decoder on the location and timing of loss events~\cite{gu2024optimizingquantumerrorcorrection,baranes2026leveraging,liu2026achieving}. See Appendix~\ref{App:BondLoss} for more details on the simulation and decoding of bond loss.

In this work, we exploit two key architectural features: photon loss is heralded at measurement, and each photon interacts with only a bounded number of atoms before being measured. Following Ref.~\cite{gu2024optimizingquantumerrorcorrection}, we partition each photon trajectory into intervals between successive CZ gates and assign a conditional probability that the photon was lost in each interval. Conditioned on loss in a given interval, all later CZ gates involving that photon are replaced by correlated Pauli $Z$ errors on the corresponding neighboring atoms, consistent with the loss-propagation analysis of Refs.~\cite{yu2025processingdecodingrydbergdecay,yu2026locating}. 
This replacement can be understood as an effective Pauli representation of the uncertainty induced by photon loss. Since photon loss can be treated as an unobserved projection, the photon state after the loss event is unknown. If the photon would subsequently have participated in CZ gates, this unknown state propagates through those gates as correlated $Z$-type uncertainty on the neighboring atoms. For details, see Appendix~\ref{App:BondLoss} and Fig.~\ref{fig:CondProbCZ}.

This construction yields a detector error model in which loss-induced correlations are represented by conditional correlated $Z$ errors. Because each photon has bounded degree, the induced errors remain spatially and temporally local, avoiding the severe effective-distance degradation associated with more pessimistic loss approximations. Numerically, we sample heralded loss configurations and, for each configuration, sample the associated conditional correlated-$Z$ error.
As shown in Fig.~\ref{fig:FittingMemory}, this loss-aware model preserves optimal distance scaling in our simulations.

\subsection{FT memory channel simulations}\label{FTsim}
We begin with fault-tolerant memory-channel simulations of the RHG lattice to evaluate our loss-aware decoder under the circuit-level noise model. The logical qubit is encoded in the three-dimensional cluster state and propagated for $d$ layers. Logical failure is determined by comparing the inferred logical observable with the ideal outcome, and the logical error rate (LER) is estimated as a function of the physical loss probability and code distance $d$.

The RHG lattice is bipartite, and therefore naturally matches the photon--atom structure of our hardware: one partite set can be assigned to photons and the other to atoms, as shown in Fig.~\ref{RHGconstruct}. In this placement, the two complementary logical correlation surfaces can be chosen to be supported on different physical subsystems, one on photons and the other on atoms. Throughout this section and Sec.~\ref{sec:CliffordGates}, we report results for the bipartite generation scheme described in Sec.~\ref{sec:GS} and illustrated in Fig.~\ref{fig: bipartite-scheme}.

We next use the circuit-level error model summarized in Table~\ref{tab:error_model}. All cavity-mediated operations and photon detection are assigned a single photon-loss parameter $p$. For cavity-mediated operations, $p$ represents an effective loss probability combining the cavity efficiency $\eta$ of Eq.~\ref{eq: effiency of vprep}, averaged over different gates, with optical path losses between consecutive gates. Photon-detection loss is set to the same value for simplicity. Since atomic measurement consists of STAP followed by photon detection, we assign it a loss probability of $2p$.

\begin{table}[ht]
    \centering
    \begin{tabular}{lcc}
        \hline
        & Photon & Atom \\
        \hline
        Initialization      & $p$   & $0$        \\
        CZ (photon–atom)    & $p$   & $0$        \\
        STAP              & --    & $p$        \\
        Measurement         & $p$   & $2p$ \\
        \hline
        \end{tabular}
    \caption{Error model used in the threshold simulations. We focus solely on photon loss as this is the most probable error mechanism. For simplicity, all gates are modeled with a single photon loss parameter $p$, except atom measurement as it includes both STAP and photon detection.}
    \label{tab:error_model}
\end{table}

\begin{figure}[!t]
\centering
    \subfloat{
    \stackinset{l}{2pt}{t}{2pt}{(a)}{
    \includegraphics[width=0.95\linewidth]{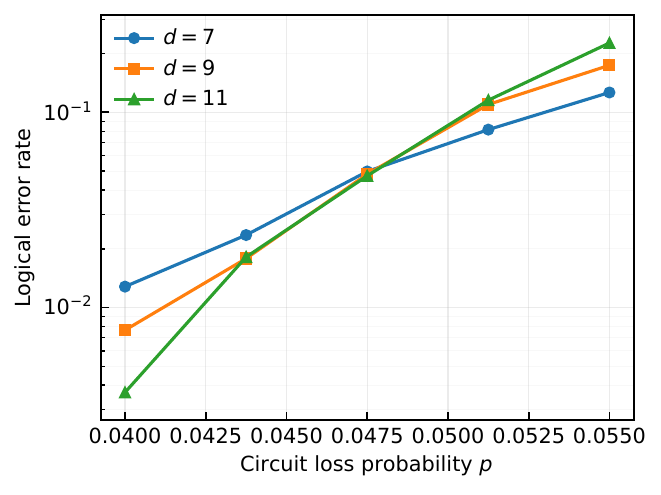}}
    }
    \\
    \subfloat{
    \stackinset{l}{2pt}{t}{2pt}{(b)}{
    \includegraphics[width=0.95\linewidth]{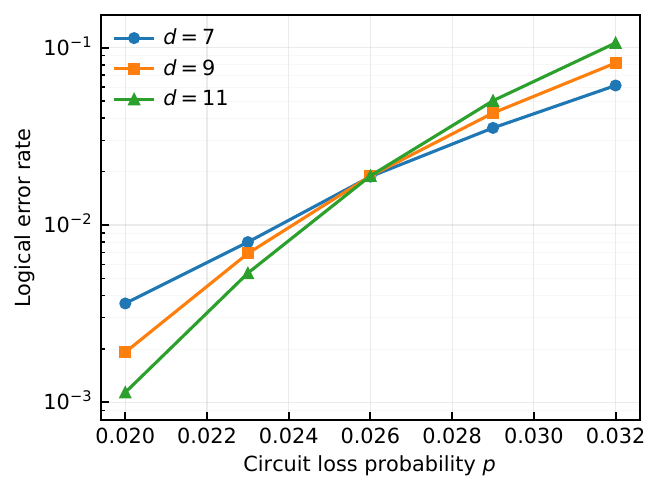}}} 
    \\
    \subfloat{\stackinset{l}{2pt}{t}{2pt}{(c)}{
    \includegraphics[width=0.95\linewidth]{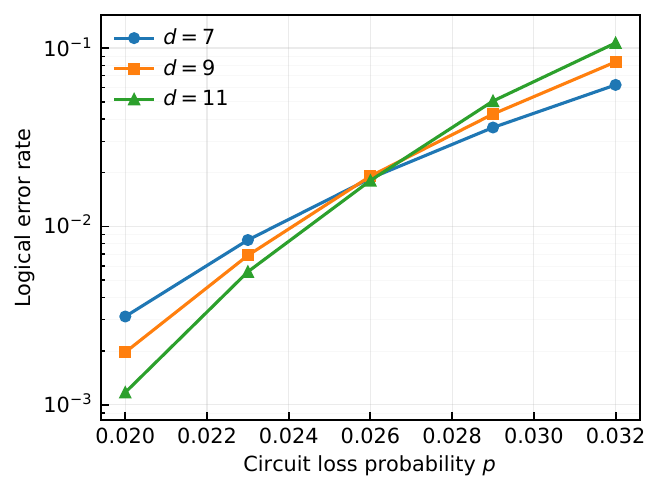}}}
    \caption{Memory simulation of the RHG lattice under the circuit-level loss model of Table~\ref{tab:error_model} for (a) logical $X$ with a photon-supported correlation surface, (b) logical $X$ with an atom-supported correlation surface, and (c) logical $Y$, whose correlation surface spans both atomic and photonic qubits. The photon-supported case has a higher threshold, consistent with an effectively iid loss process over the six lossy operations of each bulk photon. The atom-supported case is dominated by photon-loss-induced bond loss during photon--atom interactions, giving a threshold of approximately $2.6\%$ per physical gate. Near this threshold, the logical $Y$ simulation is dominated by the atomic component and closely matches the atom-supported logical $X$ case. The y-axis is logarithmic to make low-error differences visible.}
\label{Fig:circuit_loss}
\end{figure}

We first consider a logical correlation surface supported entirely on photonic qubits. Since atoms are assumed not to be lost during CZ gates, photon loss does not induce bond-loss propagation on this surface. Loss events are therefore effectively iid-like, with probabilities determined by the number of lossy operations involving each photon. A bulk photon undergoes six lossy operations: initialization, four CZ gates, and measurement. Thus, the effective loss probability is
$p_{\mathrm{eff}} = 1-(1-p)^6 = 6p - 15p^2 + O(p^3)$. 
Equating this with the RHG iid loss threshold of approximately $25\%$ gives $p \simeq 4.7\%$, consistent with the threshold observed in Fig.~\ref{Fig:circuit_loss}(a).

The complementary case is a correlation surface supported entirely on atomic qubits. Here, errors arise mainly from photon-loss-induced bond loss, with an additional contribution from measurement loss, yielding a lower threshold of $\sim2.6\%$ per physical gate, as shown in Fig.~\ref{Fig:circuit_loss}(b). Although photons are lossy and atoms are not, the photonic correlation surface performs better because photon loss is directly heralded, whereas the induced bond-loss location is not. Moreover, a photon’s total loss probability scales with its number of operations, while the effective bond-loss probability seen by an atom scales with the operations of its neighboring photons, which is larger.

For a general logical state, the correlation surface spans both atomic and photonic qubits. To probe this case, we simulate logical $Y$-state memory. Since the photonic threshold is substantially higher, logical errors from the photonic part of the correlation surface are negligible at the atomic threshold. Consequently, both the overall threshold and the LER are dominated by the atomic component, as shown in Fig.~\ref{Fig:circuit_loss}(c).

The photon-loss threshold of $\sim2.6\%$ per physical gate is comparable to the predicted $\sim97$--$98\%$ efficiencies of the protocols presented in Sec.~\ref{subsec: Physical layer - Unit-Cell}. However, those efficiency estimates do not include optical path loss and finite-fidelity effects, and practical fault-tolerant computation requires operation below threshold. These considerations motivate further work on optimized hardware protocols, improved decoding strategies, and error-correction schemes better matched to the hardware noise model.

\section{Fault-tolerant quantum gates}
\subsection{Clifford gates}\label{sec:CliffordGates}

Clifford gates map Pauli operators to Pauli operators under conjugation and are therefore central to stabilizer-based quantum error correction. A standard route to fault-tolerant logical Clifford gates is transversality, where each physical qubit in a code block undergoes an independent single-qubit operation or interacts only with the corresponding qubit of another block. This locality prevents single faults from spreading within a code block. However, not every code admits transversal implementations of the full Clifford group, since this depends on the code symmetries, stabilizers, and boundary conditions. In the surface code, Clifford operations can be implemented in an almost transversal manner: $X$, $Z$, and CNOT are realized through local or pairwise interactions, while Hadamard ($H$) and phase ($S$) require folded or geometric constructions, effectively acting transversally on a rotated or mirrored code~\cite{horsman2012surface, moussa2016transversal, webster2023transversal, wan2024iterative, breuckmann2024fold, chen2026transversal}. We extend these ideas to the RHG cluster-state architecture, where the photon-mediated nonlocal connectivity of compound photon--atom hardware enables logical $H$, $S$, and $\mathrm{CNOT}$ within MBQC. As detailed in Appendix \ref{App:CliffordGates}, all three gates are realized through gate-specific variants of a unified connectivity-based strategy.

\begin{figure*}
    \centering
\includegraphics[width=0.75\textwidth]{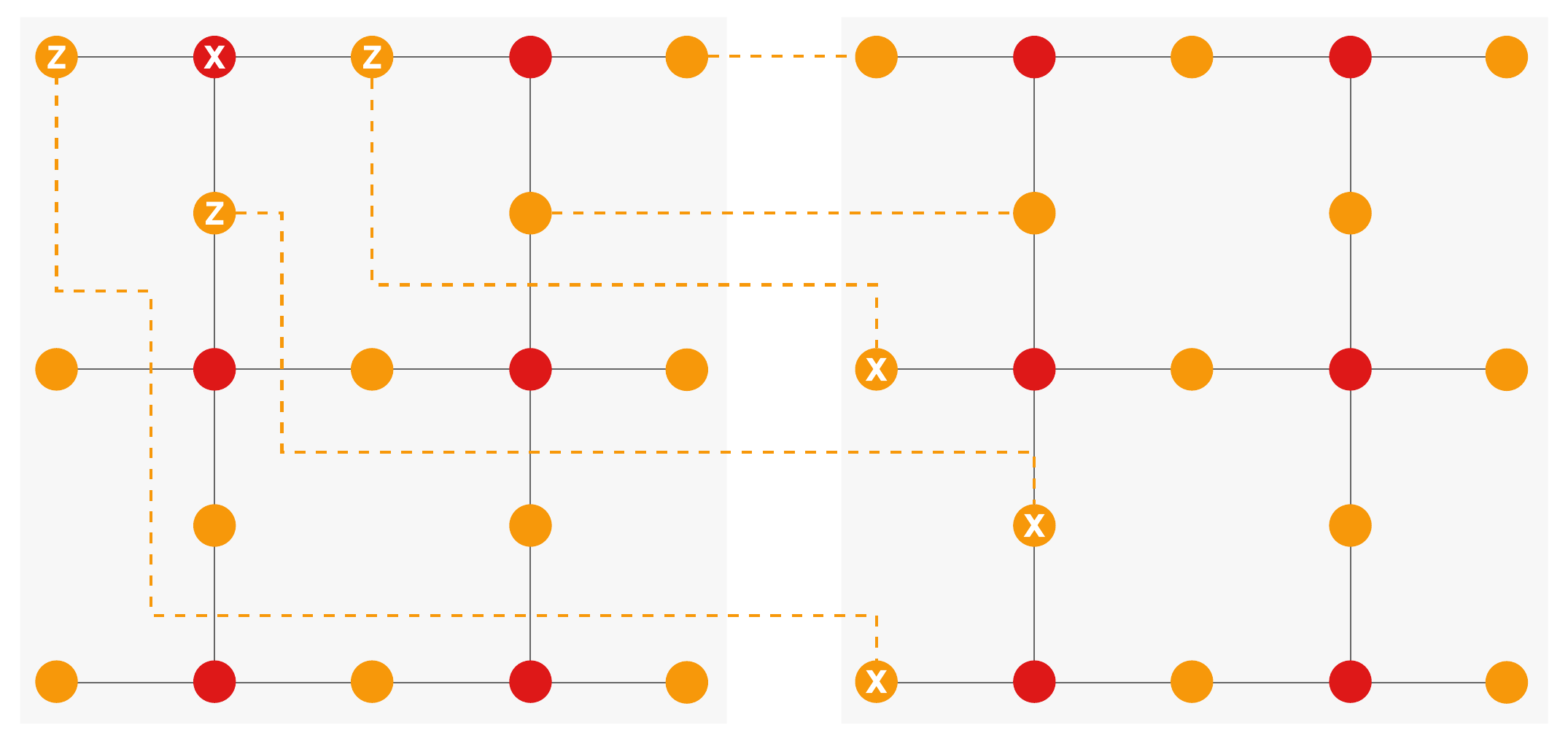}
\caption{Implementation of the logical Hadamard gate $H$ on the RHG lattice, illustrated for $d=3$. Two consecutive dual layers are connected by CZ gates with a $90^\circ$-rotated pattern, mapping rows of the first layer to columns of the second. Dashed orange curves indicate the rotated CZ connections. The white $Z$ operators acquired on qubits in the first layer by the corresponding data qubits of the second layer are canceled when the associated $X$-basis measurements are multiplied together with the red SL qubit measurement in the first layer, forming a valid check. The rotated connectivity exchanges the logical $X_L$ and $Z_L$ correlation surfaces and implements a domain wall between dual and primal check types.}
\label{Hgate}
\end{figure*}

The logical Hadamard gate, shown in Fig.~\ref{Hgate}, is implemented by introducing a domain wall in the RHG lattice. Instead of inserting a primal layer after a dual layer, we insert a second dual layer with CZ connections rotated by $90^{\circ}$. This exchanges the logical $X_L$ and $Z_L$ correlation surfaces, realizing the logical Hadamard without additional qubits or gates beyond those of the memory construction. Consequently, its logical error rate is essentially indistinguishable from that of the identity channel~\cite{bombin2023logical}.

\begin{figure}
    \centering
\includegraphics[width = 0.9\linewidth]{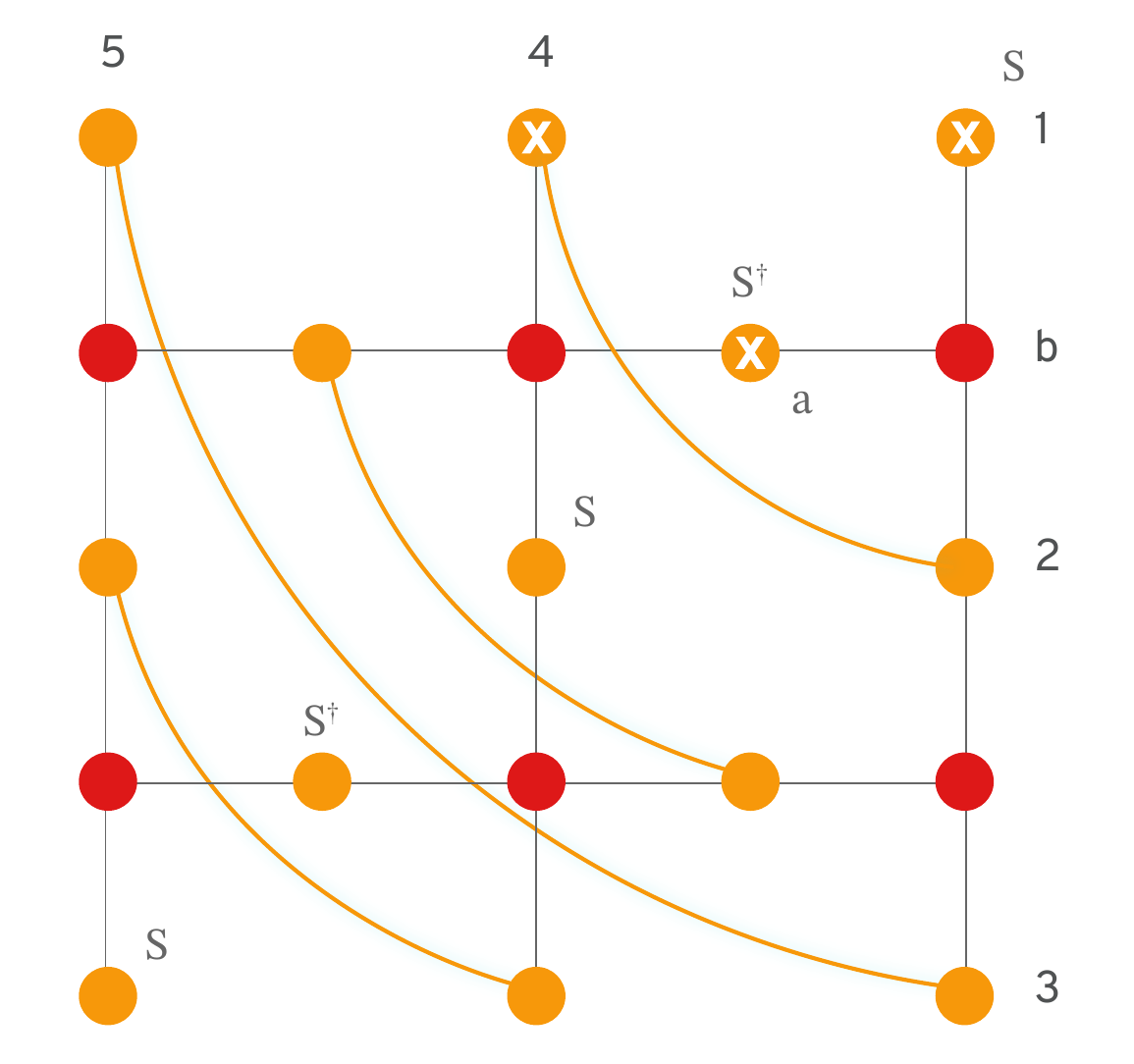}
\caption{Implementation of the logical phase gate $S$ on the RHG lattice, illustrated for the transformation $\ket{+}_L\to\ket{y^+}_L$ at distance $d=3$. Data qubits in the primal layer are connected to their mirror partners across the diagonal by additional CZ gates, shown as orange curved lines. Qubits on the diagonal are measured in alternating $S$ and $S^\dagger$ bases. The white $X$-basis measurements indicate an example check, which is extended after the gate by including the stabilizer centered at the qubit $b$.}
\label{Str}
\end{figure}

The logical phase gate $S$, shown in Fig.~\ref{Str}, is implemented by a fold-transversal construction. Data qubits in the primal layer are connected by CZ gates to their mirror images across the diagonal, while diagonal qubits are measured in an alternating $Y^+$ or $Y^-$ basis. This implements $X_L \to Y_L = i X_L Z_L$. On the gate layer, primal checks are redefined by incorporating SL qubits from the dual checks, which cancel the additional $Z$ contributions acquired by primal data qubits from the CZ and $S$ operations. The dual checks remain unchanged. Because the redefined checks include additional SL qubits, the resulting syndrome graph contains hyperedges that are not directly compatible with MWPM. Following Refs.~\cite{cain2025fast, turner2025scalable}, these hyperedges can be decomposed into matching-compatible subgraphs, enabling efficient MWPM decoding. See Appendix~\ref{App:Sim_details} for details.

Compared with state injection, which is non-fault-tolerant and requires either doubling the number of qubits per layer or adding $O(d)$ time steps~\cite{bombin2023logical, gidney2024inplace}, the fold-transversal construction uses only one additional CZ gate per data qubit in a single layer, with no extra qubits or time steps. The main hardware constraint is that the diagonal CZ gates couple same-type qubits, conflicting with the bipartite structure of our architecture. This can be resolved either by using a mediating photon or atom to realize the effective entangling operation, or by using the STAP gate to convert half of the atoms to photons.

\begin{figure*}
    \centering
\includegraphics[width=0.6\linewidth]{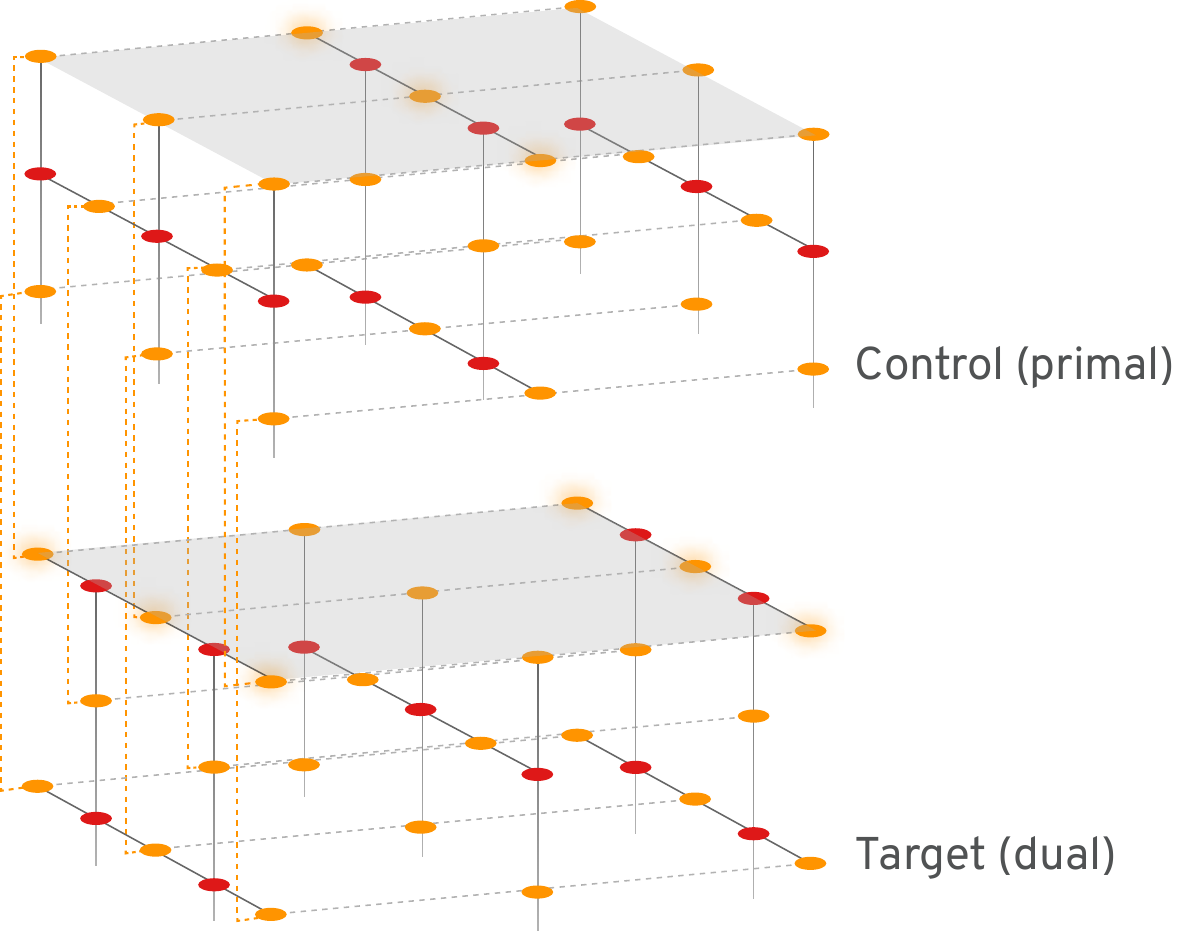}
    \caption{Transversal implementation of the logical CNOT gate on the RHG lattice. The operation is realized by pairwise physical CZ gates, shown as dashed orange lines, between the primal layer of the control logical qubit and the dual layer of the target qubit. For clarity, only a subset of the two RHG blocks is shown, and the gate is depicted in the first layers, corresponding to logical Bell-pair generation. The same construction can also be applied to existing logical qubits. In the figure, the target $Z_L$ correlation surface acquires an additional $Z$ string on the primal layer of the control, which propagates through the structure and cancels when continued through the successive dual layers of the control block. The resulting Bell-pair correlation surface, $Z^{c}_{L}Z^{t}_{L}$, is shown in light gray, with the glowing qubits indicating its support.}
    \label{CNOTgate}
\end{figure*}

The logical CNOT, shown in Fig.~\ref{CNOTgate}, is implemented transversally by applying pairwise CZ gates between the primal layer of the control block and the dual layer of the target block. This realizes $X_L^c \to X_L^c X_L^t$ and $Z_L^t \to Z_L^c Z_L^t$, while leaving $Z_L^c$ and $X_L^t$ invariant. The checks on the gate layers are redefined by incorporating SL qubits from the opposite check sublattice in the other block, which cancel the additional $Z$ contributions acquired by the data qubits. Compared with lattice surgery, which requires an ancilla logical patch of size $O(d^2)$ and $O(d)$ syndrome-measurement rounds per joint measurement~\cite{bombin2023logical}, this implementation uses a single layer of pairwise CZ gates between the two existing logical blocks, with no additional logical qubit and only $O(1)$ temporal overhead.

\begin{figure}[!t]
\centering
    \subfloat{
    \stackinset{l}{2pt}{t}{2pt}{(a)}{
    \includegraphics[width=0.95\linewidth]{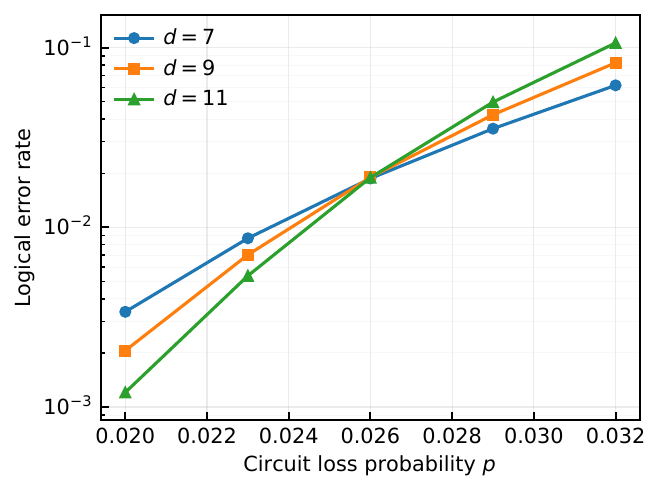}}
    }
    \\
    \subfloat{
    \stackinset{l}{2pt}{t}{2pt}{(b)}{
    \includegraphics[width=0.95\linewidth]{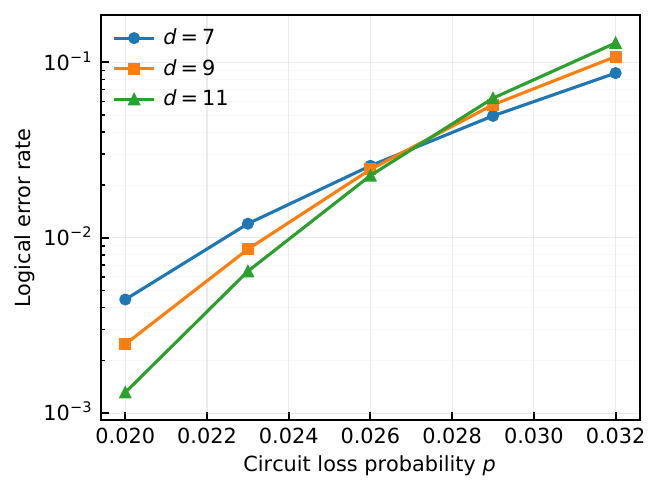}}} 
    \\
    \subfloat{\stackinset{l}{2pt}{t}{2pt}{(c)}{
    \includegraphics[width=0.95\linewidth]{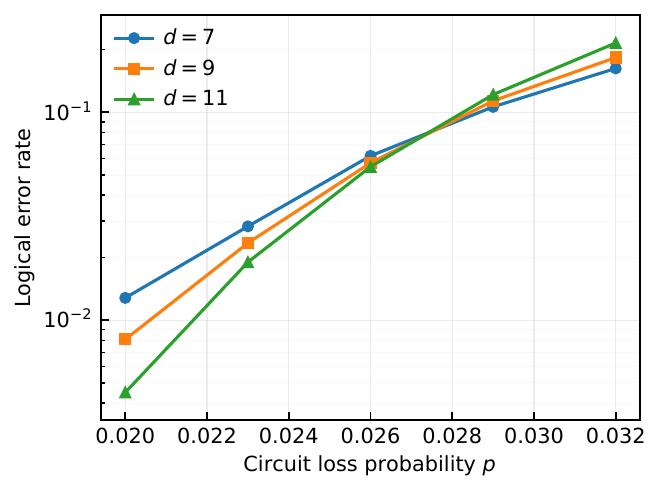}}}
    \caption{Logical-channel simulations on the RHG lattice under the circuit-level loss model of Table~\ref{tab:error_model} for (a) the Hadamard gate $H$, (b) the phase gate $S$, and (c) the CNOT gate. The gates are implemented transversally or fold-transversally and introduce only additional lossy CZ operations confined to a single layer. Their thresholds match the memory-channel threshold of Fig.~\ref{Fig:circuit_loss}, while the additional CZ operations and modified syndrome graphs increase the LER below threshold. The y-axis is logarithmic to make low-error differences visible. Simulation details are provided in Appendix~\ref{App:Sim_details}}
     \label{fig: CliffordResults}
\end{figure}

Because these gates are implemented transversally or fold-transversally, their error-correction structure closely mirrors that of the memory channel. We use the bipartite photon--atom generation scheme of Sec.~\ref{sec:GS}, illustrated in Fig.~\ref{fig: bipartite-scheme}. The only additional physical error sources are the extra CZ gates required by the logical $S$ and CNOT gates, which are confined to a single layer containing $O(d^2)$ qubits. As explained in Appendix~\ref{App:Sim_details}, using the decoding framework of Refs.~\cite{cain2025fast, turner2025scalable}, all logical Clifford gates can be decoded efficiently with MWPM. As shown in Fig.~\ref{fig: CliffordResults}, all gates achieve thresholds of $\sim2.6\%$, matching the memory-channel threshold in Fig.~\ref{Fig:circuit_loss}. Although the thresholds are unchanged, the additional CZ operations and modified syndrome graphs increase the LER below threshold. For the phase gate, this increase is about $10$--$15\%$ relative to the logical-$Y$ memory-channel baseline, while remaining well below lattice-surgery-based implementations~\cite{bombin2023logical}. The effect is more pronounced for the two-qubit CNOT channel, whose logical error rate is approximately $1.5$--$2$ times the naive estimate obtained by combining the memory-channel error rates of the two logical blocks. This is still about a factor of two lower than the expected LER of lattice-surgery-based CNOT gate based on its larger space-time volume~\cite{bombin2023logical}. In both cases, the transversal or fold-transversal implementation avoids the large additional space-time overhead associated with lattice surgery. These results confirm that the fault-tolerance properties of the RHG construction are preserved across the Clifford set.

\subsection{Magic state preparation in the MBQC framework}\label{sec:NonClifford}

Unlike physical qubits, logical qubits encoded in quantum error-correcting codes generally do not support arbitrary rotations transversally or otherwise fault-tolerantly. Logical circuits are therefore typically compiled into the Clifford+$T$ gate set, where the non-Clifford $T$ gate is implemented by gate teleportation using a high-fidelity magic state, followed by measurement-dependent Clifford corrections.

We summarize two complementary routes for preparing non-Clifford resource states in our MBQC architecture: a code-teleportation-based route and a direct MBQC adaptation of magic state cultivation. Both protocols are expressed in a foliated cluster-state representation using the native operations of the platform, namely CZ gates, $\ket{+}$ initialization, and single-qubit measurements. In both cases, the construction tracks detector sets, correlation surfaces, and conditional Pauli corrections, and evaluates the resulting protocols under circuit-level noise with postselection and decoding for a $Y$ state. The numerical values quoted below should be viewed as architecture-level benchmarks, with the full technical construction presented in forthcoming work~\cite{non_cliffords_qs}. All quoted numerical benchmarks in this section are obtained at physical error rate $p=10^{-3}$.

The first route builds on the code-teleportation protocol of Ref.~\cite{daguerre2025code}, which transfers a magic state from the distance-$3$ qRM code to the distance-$3$ Steane code. The qRM code supplies a transversal non-Clifford $T$ gate, while a transversal CNOT between the codes teleports the logical magic state to the Steane-code block. We then convert this state into a surface-code resource by combining Bell-pair growth to a distance-$5$ color code, a $T$-check gadget that filters faulty magic states~\cite{gidney2024magicstatecultivationgrowing}, color-code-to-surface-code lattice surgery~\cite{Poulsen_Nautrup_2017}, and diagonal surface-code growth~\cite{Li_2015} up to distance $11$. Flag-qubit gadgets are included in the stabilizer-extraction circuits following Ref.~\cite{Chamberland_2018}. In the present implementation, we postselect on any nontrivial syndrome or flag event, except during diagonal growth where logical-gap postselection is used~\cite{Bomb_n_2024}. The CBQC protocol is then compiled into a bipartite MBQC cluster-state form native to our architecture. As an architecture-level benchmark, this end-to-end construction yields an LER of $\sim 8\times 10^{-9}$ with an acceptance rate of $\sim 0.2\%$ for $Y$-state simulation. The main overhead is the lattice-surgery step, whose acceptance rate is $\sim 0.3\%$. We expect substantial room for optimization because this step uses only $\sim 90$ qubits per syndrome-measurement round.

The second route adapts magic state cultivation directly to MBQC~\cite{gidney2024magicstatecultivationgrowing}. We reformulate the injection, code-growth, $T$-check, and escape stages within the foliated framework. Injection is implemented using Bell-pair constructions, growth is performed by extending the foliated lattice while preserving the stabilizer structure, and the $T$-check is translated into a graph-state construction that detects both local and logical faults. The cultivation protocol achieves an LER of approximately $2\times 10^{-6}$ at distance $3$ with an acceptance rate of about $60\%$, improving to $\sim 3\times 10^{-9}$ at distance $5$ with an acceptance rate of $\sim 15\%$. The final escape stage transfers the state from the color-code region to a larger matchable code for efficient decoding. For example, converting from $d_{\mathrm{color}}=5$ to $d_{\mathrm{surface}}=7$ yields an LER of $\sim 8.6\times 10^{-9}$ with an acceptance rate of $\sim 20\%$.

These results indicate that MBQC can support both code-teleportation-based and cultivation-based preparation of non-Clifford resource states with competitive logical error rates. The two routes offer complementary tradeoffs: code teleportation is more direct and modular, while cultivation is more efficient from a topological-code perspective. The present implementations are not yet fully optimized; in particular, detector construction in dynamically changing foliated codes, gate ordering, hook-type boundary errors, acceptance rules, and photon loss remain important directions for further study.

\section{Summary}
We have presented a blueprint for a compound photon--atom architecture for fault-tolerant quantum computation, combining the long-range connectivity and scalability of photonic systems with the controllability and deterministic interactions enabled by cavity-coupled atomic qubits. The central ingredient is a near-deterministic photon--atom CZ gate implemented using cavity QED and a modified Duan--Kimble protocol, which circumvents the probabilistic entangling limitations of linear-optical photonic approaches while preserving the advantages of measurement-based quantum computation.

At the physical layer, we described a cavity-based implementation using single $^{87}$Rb atoms coupled to high-cooperativity optical resonators. The platform supports the elementary operations required for large-scale quantum computing, including single-photon generation, photon--atom entangling gates, heralded atomic state preparation and measurement via STAP. We further introduced level-engineering techniques that suppress unwanted transitions and allow improved gate schemes. The architecture naturally enables operation times on the order of tens of nanoseconds, offering substantial speed advantages over many existing platforms.

At the architectural level, we showed how the compound quantum hardware supports large-scale cluster-state generation for MBQC. Because photons provide effectively unrestricted connectivity, the architecture avoids many geometric constraints associated with local-interaction platforms. At the same time, atomic-qubit reuse significantly reduces the number of cavities and active control components. We introduced two generation schemes for the RHG lattice, bipartite and interleaved, and analyzed the tradeoff between spatial overhead and photonic delay requirements.

We further developed a fault-tolerant framework based on foliated cluster-state constructions such as the RHG lattice. A key contribution is a hardware-aware simulation framework tailored to compound photon--atom systems, explicitly incorporating asymmetric loss processes and correlated errors induced by photon loss during photon--atom interactions. This framework integrates recent ideas in loss-aware decoding and MWPM-based Clifford-gate decoding, preserving linearity at the detector-error-model level while retaining the relevant loss-induced correlations. Numerical simulations demonstrate photon-loss thresholds of $\sim 2.6\%$ per physical gate, equivalent to $\sim 15\%$ total loss over a photon trajectory, with optimal scaling of the LER below threshold.

Beyond quantum memory, we showed that the full Clifford gate set can be implemented fault-tolerantly in the RHG framework using the native connectivity of the architecture. Logical Hadamard, phase, and CNOT gates admit transversal or fold-transversal MBQC realizations and can be decoded efficiently with MWPM-based techniques. Their thresholds match the logical identity channel, with gate-dependent LER increases from the additional lossy CZ operations and modified syndrome graphs. The overhead is limited to $O(d^2)$ CZ gates and requires no additional logical qubits or RHG lattice layers. Together, these results show that RHG fault tolerance is preserved under logical Clifford operations.

Finally, we investigated two complementary approaches for non-Clifford resource-state preparation in MBQC: code teleportation and magic state cultivation. By reformulating these protocols within a foliated cluster-state framework, we demonstrated that MBQC can support high-fidelity preparation of logical non-Clifford states while maintaining compatibility with realistic decoding and hardware constraints. Numerical results indicate strong logical-error suppression with scaling consistent with the expected distance dependence.

Overall, the architecture presented here provides a concrete pathway toward scalable, fault-tolerant quantum computation based on compound photon--atom interactions. While substantial experimental and engineering challenges remain, including large-scale optical integration, improved cavity fabrication, switching infrastructure, and full system-level optimization, the combination of deterministic entangling operations, fast timescales, flexible connectivity, and native compatibility with MBQC makes this approach a compelling candidate for large-scale quantum computing. Looking ahead, the same long-range connectivity may enable efficient implementations of qLDPC codes~\cite{vasic2025quantum,panteleev2022asymptotically,leverrier2022quantum,lin2022good,bravyi2024high} with little additional experimental overhead. The combination of fast photonic gates and short-term atomic memory may also provide a useful timing degree of freedom, allowing future implementations to balance total runtime against the required switching, control, and classical-processing rates. The framework developed in this work establishes both the physical and fault-tolerant foundations for future generations of compound photonic--atomic quantum processors.

\begin{acknowledgments}
We thank Juval Bechar, Omri Davidson, Itamar Sela, and Nimrod Shenker for fruitful discussions and valuable comments on the manuscript.
B.D. holds the Dan Lebas and Roth Sonnewend Professorial Chair of Physics.
\end{acknowledgments}

\appendix

\section{Simulation details for RHG memory and Clifford channels}
\label{App:Sim_details}

All error-correction simulations are performed in a custom simulation environment based on \texttt{STIM}~\cite{gidney2021stim} sampling and MWPM decoding with PyMatching~\cite{higgott2021pymatchingpythonpackagedecoding, pymatchingv2, Higgott2025sparseblossom}. The simulator implements our architecture-specific loss-aware decoding procedure described in Sec.~\ref{sec: EM} and Appendix~\ref{App:BondLoss}. When shown, error bars indicate the standard error of the mean. In most cases, they are smaller than the marker size and are therefore omitted. Thresholds are estimated from the finite-size crossing points.

For the logical $S$ and CNOT simulations, where the modified detector definitions produce syndrome graphs containing hyperedges, we use a custom pre-decoder following the method of Ref.~\cite{cain2025fast}. Ref.~\cite{cain2025fast} proves that, for surface-code circuits with transversal and fold-transversal Clifford gates, each reliable logical Pauli product admits a matchable syndrome subgraph obtained by selecting same-basis checks along the back-propagation path of that product through the Clifford circuit.

For the memory-channel simulations shown in Fig.~\ref{Fig:circuit_loss}, we use the circuit-level loss model of Table~\ref{tab:error_model}. For the iid benchmark shown in Fig.~\ref{Fig: iid_loss_sim}, we instead use an iid measurement-loss model. In both cases, the RHG lattice contains $t=d$ noisy layers, where $d$ is the code distance. Logical-$Y$ memory is simulated using two additional capping layers, giving a total of $d+2$ layers. The first and last layers are treated as noiseless boundary layers, and the loss model is applied only to the central $d$ layers. These caps implement the logical-$S$ boundary transformation, corresponding to perfect injection and measurement, so that only the memory channel itself is simulated. Each data point uses at least $3000$ loss realizations and $10{,}000$ decoding iterations per realization.

For all Clifford-gate simulations shown in Fig.~\ref{fig: CliffordResults}, we use the circuit-level loss model of Table~\ref{tab:error_model}. For the logical-$H$ simulation shown in Fig.~\ref{fig: CliffordResults}(a), we use $t=d$ layers and no capping layers. The construction starts from an RHG lattice and implements the logical Hadamard by rotating the coordinates of the final two layers, realizing the logical mapping $X \mapsto Z$. The first-layer
data qubits are atoms. Each data point uses at least $3000$ loss realizations and $10{,}000$ decoding iterations per realization.

For the logical-$S$ simulation shown in Fig.~\ref{fig: CliffordResults}(b), we use $t=d$ layers and no capping layers. The fold-transversal $S$ operation is implemented on the third layer from each temporal boundary, corresponding to layers $2$ and $t-3$ in zero-based indexing. The first-layer
data qubits are atoms. Each data point uses at least $3000$ loss realizations and $10{,}000$ decoding iterations per realization.

For the logical-CNOT simulation shown in Fig.~\ref{fig: CliffordResults}(c), we simulate control and target RHG blocks with $d+2$ and $d+3$ total layers, respectively. The first two layers of both blocks and the final target layer are treated as noiseless boundary layers, leaving exactly $d$ noisy layers in each block. Using one-based indexing, transversal CNOT connections are applied between layer $d$ of the two blocks, so the CNOT-connection layer is noisy. Logical $S$ operations are applied as boundary transformations on the first control layer and final target layer. The control block starts from a primal layer and the target block from a dual layer, with opposite photon--atom assignments: the first-layer data bipartition is assigned to atoms in the control block and to photons in the target block. With these choices, the initial stabilizer $Y_cZ_t$ is propagated through the CNOT to the measured stabilizer $X_cY_t$. These stabilizers were chosen to yield a nontrivial correlation surface. Each data point uses at least $1000$ loss realizations and $10{,}000$ decoding iterations per realization.

\section{RHG lattice as fault-tolerant memory channel} \label{App:RHG-general}
\subsection{Construction} \label{App:Construction}

The RHG lattice, introduced by Raussendorf, Harrington, and Goyal, was the first three-dimensional fault-tolerant topological quantum code supporting a universal gate set~\cite{raussendorf2007topological}. Fig.~\ref{RHGconstruct} shows an example with $d=3$ and $t=3$. The parameter $d$ is the spatial linear size of the lattice and sets the error-correction distance, defined as the minimum weight of a nontrivial logical operator. The parameter $t$ is the number of spatial layers stacked along the time direction. All qubits are initialized in $\ket{+}$, and CZ gates are applied according to the lattice connectivity, with dotted and solid lines denoting inter- and intra-layer connections, respectively. The resulting graph is an FCC-like three-dimensional cluster state. Logical information is propagated by measuring all qubits in the $X$ basis, except for the $Y$- and magic state measurements required for the phase $S$ and non-Clifford $T$ gates. See Sec.~\ref{sec:CliffordGates} and Sec.~\ref{sec:NonClifford}. For error-correction distance $d$, up to $(d-1)/2$ errors can be corrected. Each spatial layer contains $d^{2} + (d-1)^2 + d(d - 1)$ physical qubits. Throughout this Appendix, $d$ is assumed odd unless stated otherwise.

We call layers with rough top and bottom boundaries \textit{primal} layers. These are the odd layers in Fig.~\ref{RHGconstruct}, taking time to run from left to right, and are characterized by top and bottom boundaries containing $d$ qubits. The remaining layers are \textit{dual} layers, whose top and bottom boundaries contain $2d - 1$ qubits; see also Fig.~4 of Ref.~\cite{herr2018lattice}. We call qubits connected to the next layer in time \textit{data} qubits, as they propagate logical information forward. Qubits with only intra-layer connections are called \textit{syndrome-like} (SL) qubits. 

\subsection{Foliation}\label{App:Foliation}

\subsubsection{Single-qubit teleportation}
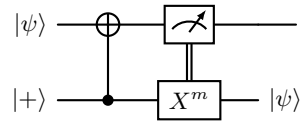
\begin{figure}
    \centering
    \begin{quantikz}
        \lstick{$\ket{\psi}$} & \targ{}    & \meter{} \wire[d][1]{c} &        & \\
        \lstick{$\ket{+}$}    & \ctrl{-1}  & \gate{X^{m}}            & \rstick{$\ket{\psi}$}
    \end{quantikz}
    \caption{Single-qubit teleportation circuit. The input state $\ket{\psi}$ is entangled with a $\ket{+}$ ancilla using a CNOT gate. Measuring the input qubit in the $Z$ basis teleports the state to the ancilla, up to the measurement-dependent Pauli correction $X^{m}$, where $m \in \{0,1\}$.}
    \label{1Qtel}
\end{figure}

A key feature of MBQC is that quantum information can be teleported using only single-qubit measurements~\cite{fujii2015quantum}. The basic circuit is shown in Fig.~\ref{1Qtel}: a state $\ket{\psi}$ is teleported to an ancilla prepared in $\ket{+}$ by entangling the two qubits with a CNOT gate and then measuring the input qubit in the $Z$ basis. Depending on the measurement outcome, a Pauli-$X$ correction may be required. This construction naturally generalizes to the teleportation of single- and two-qubit gates.

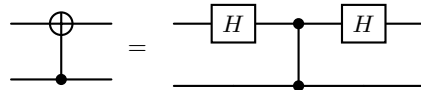
\begin{figure}
    \centering
    \begin{quantikz}
        \lstick{} & \targ{} & \\
        \lstick{} & \ctrl{-1} &
    \end{quantikz}
    =
    \begin{quantikz}
        \lstick{} & \gate{H} & \ctrl{1} & \gate{H} & \\
        \lstick{} && \control{} &&
    \end{quantikz}
    \caption{Circuit equivalence between CNOT and CZ gates.}
    \label{CZNOT}
\end{figure}

Since our construction uses only CZ gates, we replace the CNOT with its circuit equivalent: a CZ gate conjugated by Hadamard gates on the target qubit, as shown in Fig.~\ref{CZNOT}.

This substitution has a natural interpretation in the layered cluster-state picture. Data qubits in the first, dual, layer are prepared in $\ket{0}$, while those in the next, primal, layer are prepared in $\ket{+}$. Applying $H$ to the first layer maps these qubits to $\ket{+}$, after which CZ gates are applied between adjacent layers. Measuring the first layer in the $X$ basis is then equivalent to a $Z$-basis measurement preceded by $H$, which teleports the encoded information forward. Thus, a CZ gate between a $\ket{+}$ state and an arbitrary state $\ket{\psi}$, followed by an $X$-basis measurement of $\ket{\psi}$, implements single-qubit teleportation of $H\ket{\psi}$.

\subsubsection{Surface-code foliation}
\label{sc fol}
\begin{figure}
    \centering

    \subfloat{
    \stackinset{l}{2pt}{t}{2pt}{(a)}{
    \includegraphics[width=0.9\linewidth]{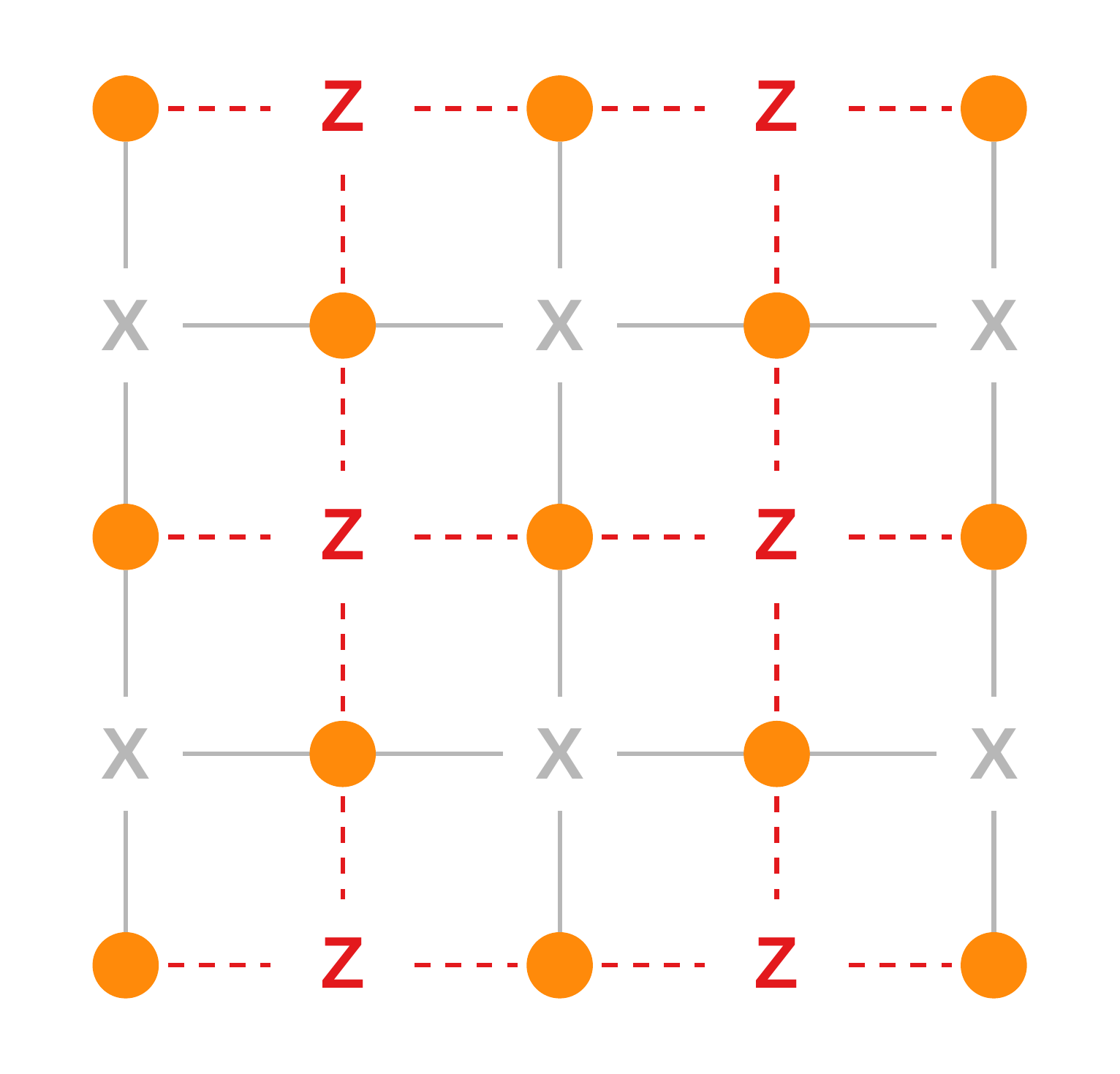}}
    }
    \\
    \subfloat{
    \stackinset{l}{2pt}{t}{2pt}{(b)}{
    \includegraphics[width=0.9\linewidth]{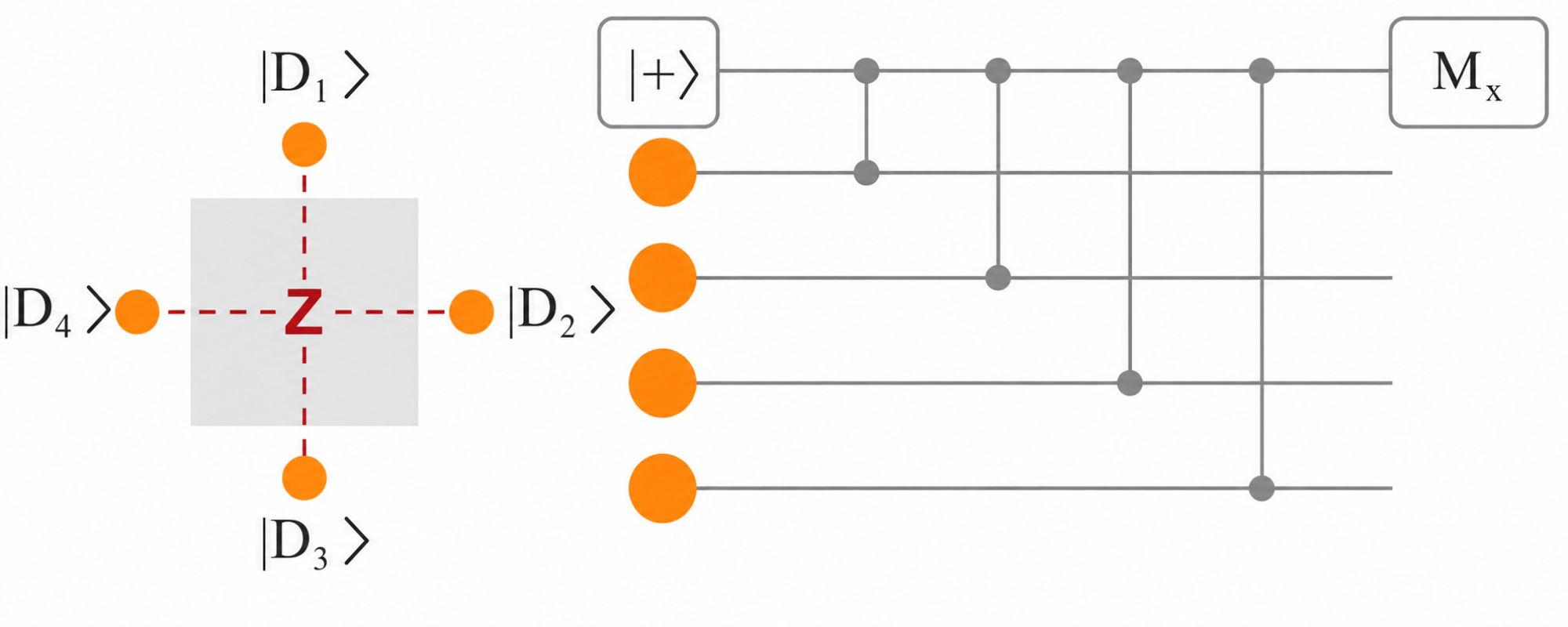}}} 
    \\
    \subfloat{\stackinset{l}{2pt}{t}{2pt}{(c)}{
    \includegraphics[width=0.9\linewidth]{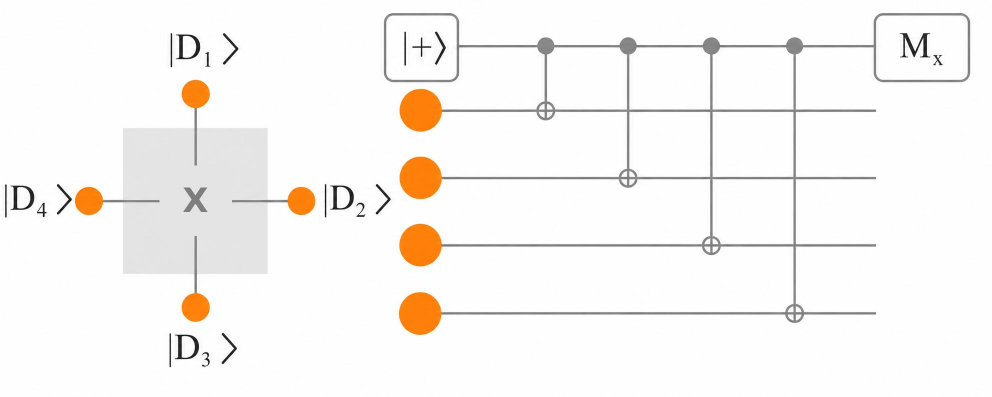}}}

    \caption{(a) The surface code on a square lattice. (b, c) $Z$- and $X$-type stabilizer measurement circuits, respectively.}
    \label{SC1}
\end{figure}

The canonical surface-code construction defines the code space using two stabilizer types, $X$-type and $Z$-type, whose parities are measured by ancilla qubits coupled to neighboring data qubits. An example is shown in Fig.~\ref{SC1}(a), where orange circles denote data qubits, and the $X$- and $Z$-type stabilizers are indicated schematically by gray X's and red Z's~\cite{fowler2012surface}. The vertical and horizontal directions have different boundary types, which later determine the orientation of the logical operators and correlation surfaces. Crucially, the two stabilizer-measurement stages correspond directly to the dual and primal RHG layers.

The correspondence between the surface code and the RHG lattice is obtained by viewing a dual layer and the following primal layer as one unit cell. All qubits in the dual layer are initialized in $\ket{+}$. The CZ pattern between SL and data qubits then acts effectively as a CNOT layer, with the data qubits serving as controls, as shown in Figs.~\ref{CZNOT} and~\ref{SC1}(b). Measuring the SL qubits in the $X$ basis completes the extraction of the $Z$-type stabilizers. CZ gates between adjacent layers then teleport each data-qubit state to its counterpart in the primal layer, together with an additional Hadamard operation. A CZ pattern between SL and data qubits in the primal layer, followed by $X$-basis measurements of the SL qubits, is therefore equivalent to measuring the $X$-type stabilizers of the \textit{same} data qubits. These two stages, first in the dual layer and then in the primal layer, constitute one complete round of syndrome measurements in the circuit-based surface-code picture.

More generally, any two-dimensional CSS code can be foliated into a three-dimensional cluster state~\cite{bolt2016foliated}, where formal definitions of CSS codes are given in Refs.~\cite{calderbank1996good, steane1996simple}. This is achieved by converting the 2D code into a cluster-state representation and stacking alternating dual and primal layers, each corresponding to one stabilizer type. The construction applies naturally to CSS codes because their $X$- and $Z$-type stabilizers are distinct. The RHG lattice is therefore the surface-code instance of this broader foliation procedure. Nevertheless, it can also be understood directly from the cluster-state perspective, without assuming prior familiarity with the surface code. This viewpoint is developed in the next section.

\subsection{RHG as a cluster state} \label{App:RHGcs}
The previous section related the RHG lattice to a foliated surface code, explaining the motivation for the construction and how the code arises from stacked surface-code layers. We now describe the same object directly as a three-dimensional topological cluster state~\cite{fowler2008topological}. This complementary viewpoint is useful because the propagation of logical information, the role of single-qubit measurements, and the structure of the error-correction checks are most naturally expressed in cluster-state language.

A cluster state is defined by its construction: all qubits are initialized in $\ket{+}$, and CZ gates are applied according to the desired graph connectivity. The resulting state has, for every qubit $i$, a stabilizer
\begin{align}
S_{i} = X_{i} \prod_{j \in N(i)} Z_{j},
\label{eq: cluster_state_stabilizer}
\end{align}
where the product runs over all qubits $j$ connected to $i$ by a CZ gate. Graphically, qubits are vertices and CZ connections are edges, so directly connected qubits are nearest neighbors in the cluster-state graph. Interpreting Fig.~\ref{RHGconstruct} in this language, each RHG qubit is the center, or $X$ component, of a stabilizer of the form in Eq.~\ref{eq: cluster_state_stabilizer}, with $Z$ operators on its neighbors. In Sec.~\ref{App:ErrorCorrection}, we explain how these local stabilizers are multiplied to construct the error-correction checks.
  
Single-qubit measurements also determine how the cluster state evolves. As discussed above, $X$-basis measurements teleport logical information forward while preserving the remaining cluster-state structure. By contrast, a $Z$-basis measurement disconnects the measured qubit from the cluster by removing its CZ bonds to neighboring qubits. This follows directly from the Gottesman--Knill stabilizer update rules, according to which stabilizers that anticommute with the measurement are removed~\cite{aaronson2004improved}. Consequently, $Z$-basis measurements are redundant in cluster-state computation: any qubit intended to be measured in the $Z$ basis can simply be omitted from the graph.

\subsection{Logical state}
\label{App:LogicalState}

To understand how logical information is stored and propagated in the RHG lattice, we distinguish between two objects: the \textit{logical operator} and the \textit{correlation surface}. A logical operator is a one-dimensional product of single-qubit Pauli operators defined on a given layer. A correlation surface is a two-dimensional product over single-qubit measurement outcomes that allows the value of this logical operator to be inferred as it propagates through the cluster. In Fig.~\ref{RHGconstruct}, the correlation surface is shown as a horizontal shaded cap. The glowing circles mark the qubits whose $X$-basis measurement outcomes contribute to the correlation surface associated with the logical operator. 

In MBQC, the correlation surface is the experimentally accessible object: by multiplying its measurement outcomes from one boundary to the other and applying the corresponding error-correction procedure, one infers the value of the logical operator. Since all RHG qubits are measured in the $X$ basis, the lattice construction must ensure that the relevant correlation surfaces consist only of single-qubit $X$ measurements.

Following the surface-code convention, we define the logical $\ket{+}_{L}$ state by starting from a \textit{primal} layer. In the foliated interpretation, the SL qubits of this layer measure the $Z$-type stabilizers of the underlying surface code, so transversal initialization of the data qubits in $\ket{+}$ mirrors the standard surface-code preparation of $\ket{+}_{L}$. In the cluster-state picture, each of the $d$ data qubits $i$ along a vertical line in the primal layer has a stabilizer containing an $X_i$ factor and $Z$ operators on its neighboring qubits. Restricting first to intra-layer contributions, multiplying these stabilizers along the line cancels the neighboring $Z$ factors and leaves a length-$d$ product of physical $X$ operators. For the initialized $\ket{+}_{L}$ state, this product has a deterministic value. It represents the initial logical operator $X_L$ and forms the starting line of the correlation surface that propagates this logical information through the cluster. As in the surface code, this deterministic surface arises only for the corresponding logical eigenstate, not for a general encoded state.

When the primal layer is connected to a dual layer, the stabilizers contributing to this vertical product acquire additional $Z$ factors on adjacent dual-layer qubits. These factors must be eliminated from any measurable correlation surface. This is achieved by sandwiching the dual layer between two primal layers, so that the additional $Z$ contributions cancel. The resulting vertical correlation surface extends across both primal layers and contains $2d$ single-qubit $X$ measurements. As logical information is propagated through successive dual-primal layer pairs, this correlation surface extends through the lattice while remaining composed solely of $X$-basis measurement outcomes.

As in the surface code, the logical operator $Z_{L}$ is defined as a horizontal string of physical $Z$ operators acting on the data qubits of the primal layer, perpendicular to the string $X_L$ and its correlation surface. This operator anti-commutes with the correlation surface associated with $X_{L}$ as they share only a single qubit, thereby flipping its value and hence the sign of the logical state --- switching $|+\rangle_{L}$ to $|-\rangle_{L}$, as expected. The corresponding $Z_{L}$ correlation surface is a horizontal string of $X$ operators in the dual layer, shown as glowing circles on the cap in Fig.~\ref{RHGconstruct}, which propagates $Z_{L}$ forward in time. Since the qubits in the primal layer are measured in the $X$-basis, the value of $Z_{L}$ is intrinsically random, reflecting the standard quantum uncertainty of a $|+\rangle_{L}$ eigenstate. To initialize the logical state $|0\rangle_{L}$ instead, one begins in the dual layer. Although all physical qubits are still prepared in $|+\rangle$, their propagation includes an additional Hadamard gate, making the primal layer effectively equivalent to initializing all physical qubits in $|0\rangle$ --- the standard $|0\rangle_{L}$ initialization of the surface code.

The same relation can be explained from the complementary dual-layer viewpoint. Suppose the memory begins with a dual layer followed by a primal layer. A vertical string of physical $Z$ operators on data qubits in the dual layer represents the logical operator $X_L$. If these dual-layer qubits were explicitly measured in the $Z$ basis, their product, together with the neighboring $X$-basis measurements in the following primal layer, would form a stabilizer constraint. Since $Z$-basis measurements are redundant in a cluster state, these qubits can instead be omitted from the graph. The logical information carried by this unrealized $Z$ string is then inferred from the measurable product of neighboring $X$-basis outcomes in the primal layer.

Thus, the logical operator can be viewed as an unmeasured Pauli string on an initial layer, while the correlation surface is the measurable product of $X$-basis outcomes that propagates its value through the cluster. Their product forms a stabilizer of the cluster state, allowing the logical observable to be inferred from the measured correlation surface after error correction~\cite{whiteside2017classical}.

\subsection{Error correction}
\label{App:ErrorCorrection}
\begin{figure}
    \centering
    \includegraphics[width=0.95\linewidth]{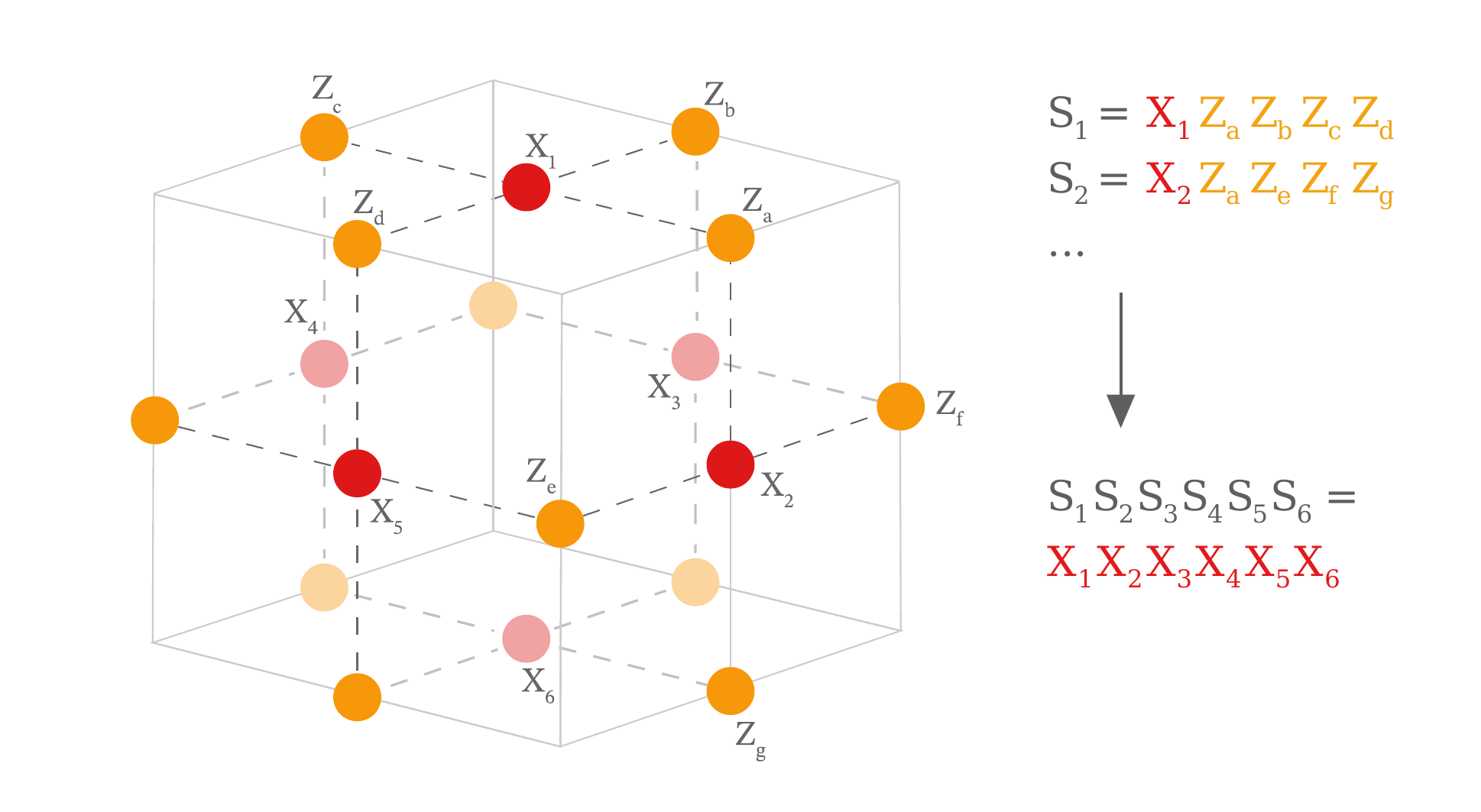}
    \caption{Each cube of the RHG lattice supports a check composed solely of single-qubit $X$ measurements, obtained by multiplying cluster-state stabilizers. Qubits that do not participate in this check belong to checks of the complementary type.}
    \label{RHGcheck}
\end{figure}

The error-correction procedure in the RHG lattice is a three-dimensional generalization of the surface-code procedure. See Ref.~\cite{fowler2008topological} for a review. The main difference is that RHG error correction uses only single-qubit measurements, all performed in the $X$ basis, rather than explicit ancilla-based stabilizer measurements. As a result, instead of mixed $X$- and $Z$-type stabilizer measurements, we construct \textit{checks}. Each check is obtained by multiplying several cluster-state stabilizers over a two-dimensional membrane enclosing a three-dimensional cube, as illustrated in Fig.~\ref{RHGcheck}.

Each bulk qubit participates in two checks, while boundary qubits can be viewed as connected to fictitious checks outside the boundary. Consequently, $Z$ errors are detected in close analogy with the surface code; see Fig.~\ref{RHGconstruct}. Not all qubits in a cube participate in a given check because the RHG lattice is a CSS-type construction with two complementary check types. Qubits absent from one check type participate in the other.

Since only $X$-basis measurements are performed, the directly detectable errors are $Z$ errors. The two check types therefore do not correspond to detecting $Z$ and $X$ errors, as in the surface code. Rather, they correspond to the two complementary RHG check sublattices, associated with the two orientations of logical correlation surfaces.

\begin{figure}
    \centering
    \resizebox{0.825\linewidth}{!}{%
        \includegraphics{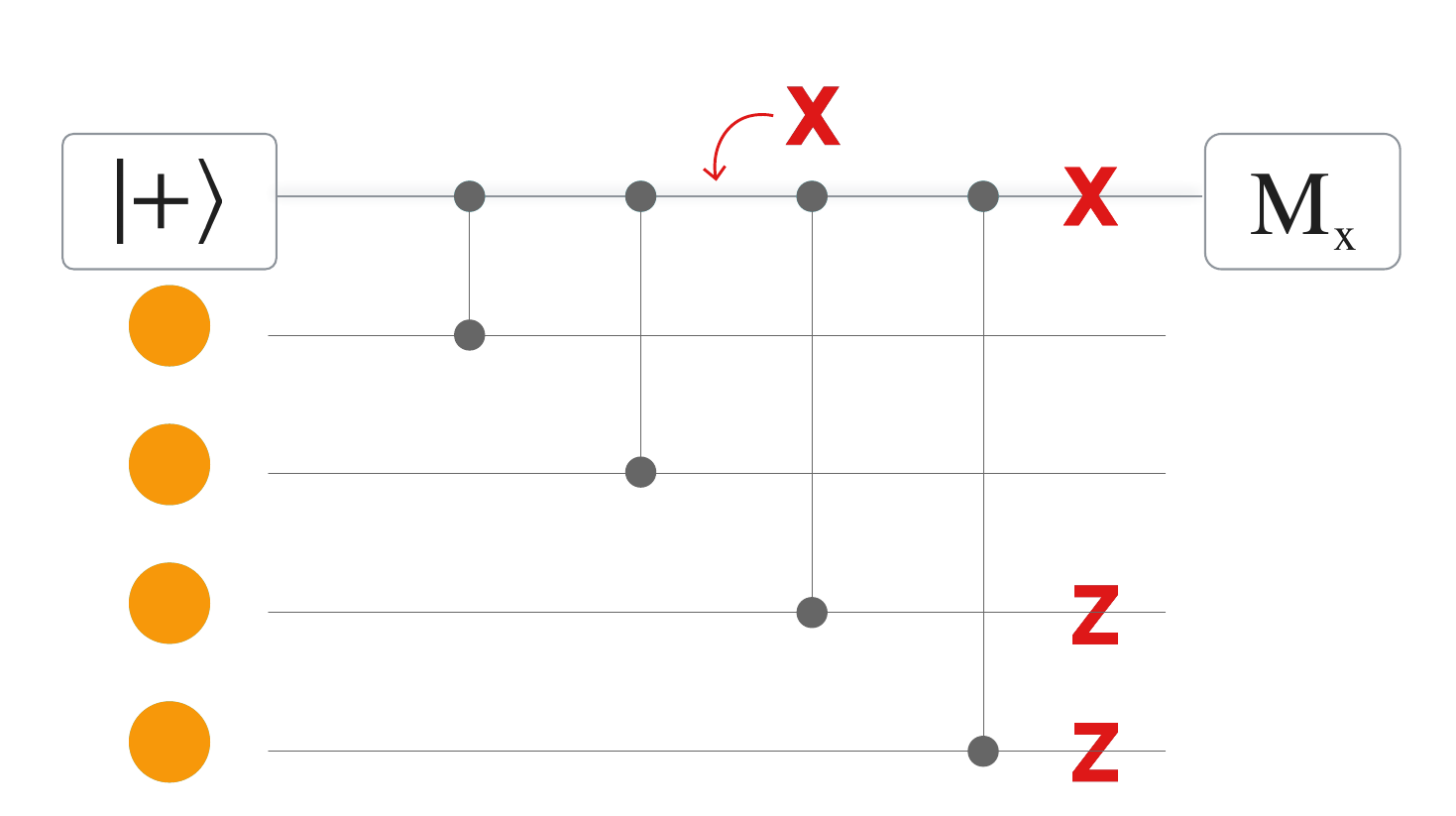}
    }
    \caption{Quantum circuit showing how an $X$ error propagates through cluster-state CZ gates into multiple $Z$ errors and an $X$ error immediately before the $X$-basis measurement, which can be ignored.}
    \label{RHGecZX}
\end{figure}

It is nevertheless sufficient to correct $Z$ errors. As explained in Ref.~\cite{fowler2008topological}, $X$ errors are either irrelevant, if they occur immediately after $\ket{+}$ initialization or just before $X$-basis measurement, or propagate through CZ gates into equivalent $Z$ errors once the qubit is entangled, as shown in Fig.~\ref{RHGecZX}.

\begin{figure}
\centering
    \stackinset{l}{2pt}{t}{2pt}{(a)}{%
      \includegraphics[width=0.95\linewidth]{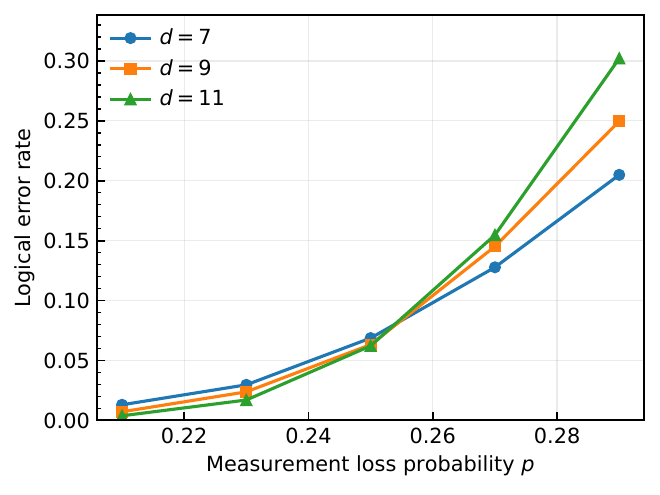}%
    }\\[0.5em]
    
    \stackinset{l}{2pt}{t}{2pt}{(b)}{%
      \includegraphics[width=0.95\linewidth]{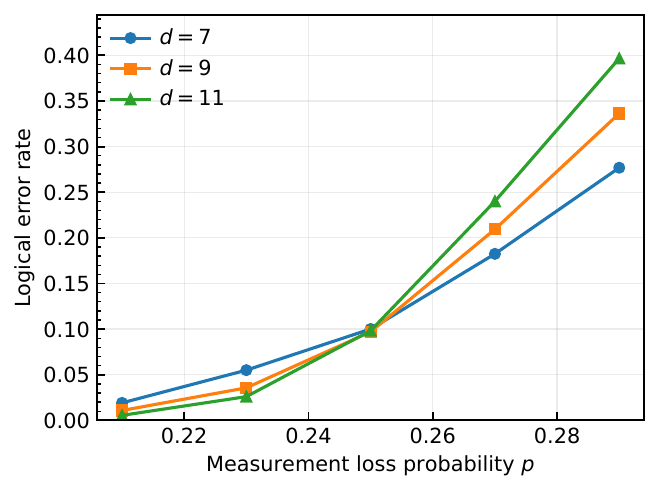}%
    }
\caption{Memory-channel simulation on the RHG lattice with an iid measurement-only loss model for (a) logical $X$, whose correlation surface is supported on one partite subgraph, and (b) logical $Y$, whose correlation surface combines the logical $X$ and $Z$ correlation surfaces. Both cases recover the known RHG loss threshold of $\sim25\%$. The LER at threshold is about $6\%$--$7\%$ for logical $X$ and roughly $1.5$ times larger for logical $Y$, reflecting the larger correlation surface and finite size effects. Unlike the hardware-inspired circuit-level model used in the main text, this iid model assigns equal measurement-loss probabilities to all qubits and serves as a generic RHG benchmark.}

\label{Fig: iid_loss_sim}
\end{figure}

Having established the check structure, we present error-correction simulation results for the generic case of iid measurement loss. We first consider a logical correlation surface supported on one partite subgraph of the RHG lattice. Independent measurement loss then reproduces the known bond-percolation threshold of approximately $0.25$~\cite{lorenz1998precise}, as shown in Fig.~\ref{Fig: iid_loss_sim}(a). Since a general logical state in the RHG lattice can have support on both complementary correlation surfaces, we also simulate the logical $Y_{L} = (\ket{0}_{L} + i\ket{1}_{L})/\sqrt{2}$ state, whose correlation surface spans both subgraphs, as shown in Fig.~\ref{Fig: iid_loss_sim}(b). The threshold is unchanged, while the larger correlation surface increases the finite-size LER by a factor of order unity, approximately $1.5$ near threshold in our simulations.

\section{Bond-loss propagation}\label{App:BondLoss}

In the main text, we distinguish between physical qubit loss and bond loss. Here we explain how physical photon loss can induce bond loss, how the resulting errors propagate to neighboring checks, and how this effect is incorporated into our detector error model.

\begin{figure}
    \centering
    \includegraphics[width=0.95\linewidth]{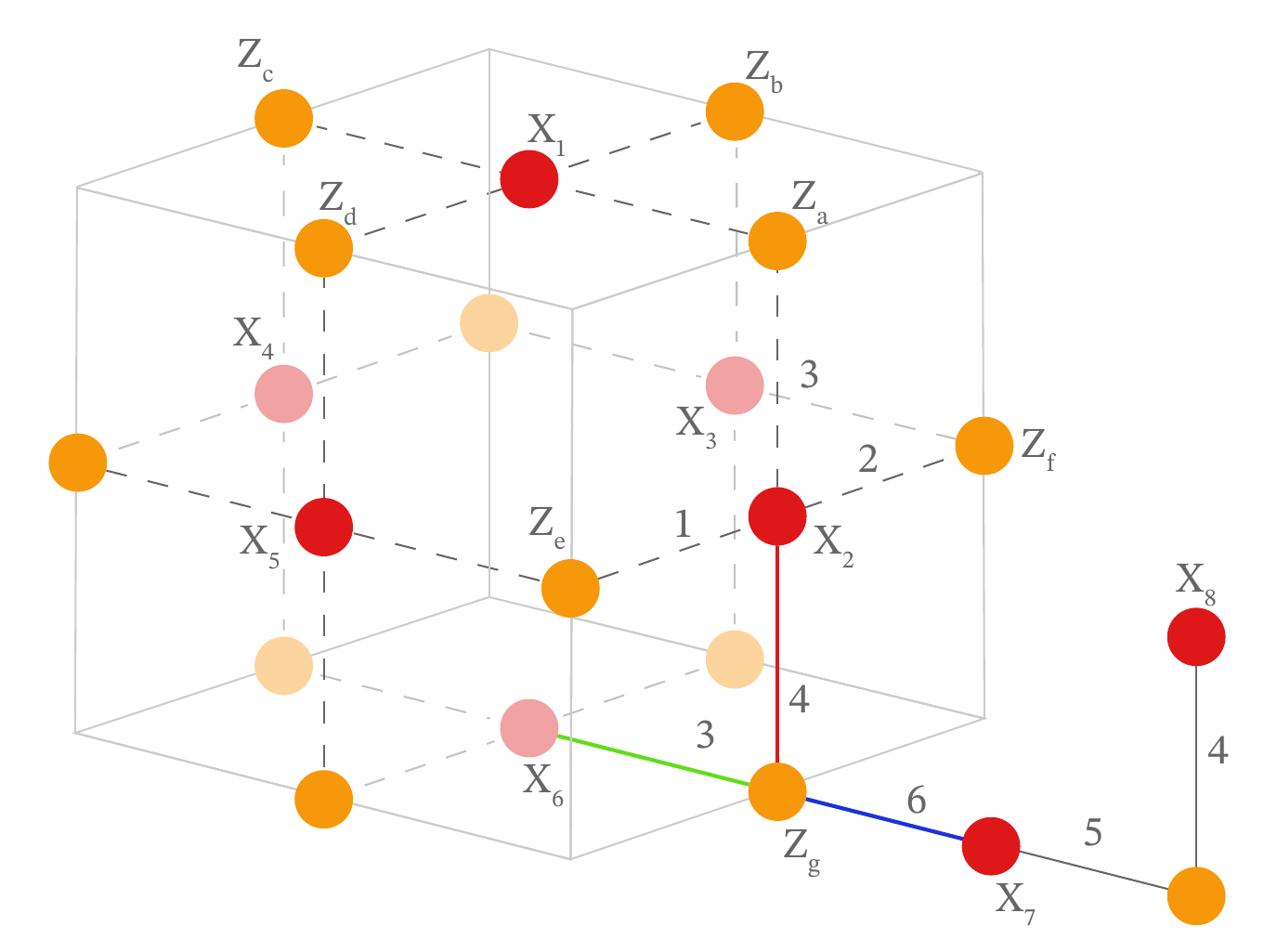}
    \caption{Time ordering of CZ operations. If qubit~$g$ is lost, the loss time is ambiguous: it may have occurred during measurement, in which case all intended bonds remain intact, or during one of the CZ gates, in which case the gates performed during and after the loss fail, removing the associated bonds and corrupting additional checks.}
    \label{fig:UnheraldedBond}
\end{figure}

In MBQC, a lost qubit or missing entangling bond can randomize the associated check. For example, if qubit~$1$ in Fig.~\ref{RHGcheck} is lost, its $X$-basis measurement outcome is unknown, and the corresponding checks are corrupted. A similar effect occurs when the qubit remains present but one of its entangling gates fails. Fig.~\ref{fig:UnheraldedBond} illustrates this mechanism in the presence of time-ordered CZ gates. Consider one complete check $C_1 = X_1X_2X_3X_4X_5X_6$, and part of a neighboring check $C_2 = X_2X_7X_8$.
Suppose that qubit~$g$ interacts with its neighbors in the order green--red--blue. If qubit~$g$ is eventually found to be lost, the effect on the checks depends on when the loss occurred. If the loss occurred only at measurement, all earlier bonds were successfully created, and the standard checks remain valid. By contrast, if the loss occurred before or during CZ$_{2,g}$, the green bond exists, while the red and blue bonds are absent. In that case, the check $C_1$ acquires a dangling operator $Z_g$ and is no longer an accessible $X$-type check because qubit~$g$ was lost. A Pauli-frame simulation that predefines the checks and then inserts single-qubit losses without modifying the bond structure would incorrectly treat $C_1$ as valid, leading to overoptimistic error-correction simulations. This is the essence of bond-loss propagation: a lost photon can corrupt neighboring checks through the absence of scheduled entangling gates, even when the neighboring qubits themselves are not physically disturbed. Without modifying the decoding process, this bond-loss information is inaccessible to the decoder, leading to degraded decoding performance in experiments.

A naive way to model this effect is to apply independent Pauli errors to every qubit that was scheduled to interact with the lost photon. Earlier approaches used fully depolarizing or fully dephasing channels on the qubits adjacent to a missing bond~\cite{whiteside2014upper,whiteside2017classical}. This is simple, but it generally overestimates the damage caused by bond loss because missing bonds do not necessarily induce independent errors on all neighboring qubits. Instead, the physical process is often better described as a correlated dephasing event acting on the set of qubits affected by the same loss event~\cite{yu2026taming}. Applying independent dephasing to affected neighbors can incorrectly randomize checks that should remain valid, whereas a correlated dephasing event preserves cancellations between added $Z$ operators when the relevant $X$-basis measurement outcomes are multiplied. Correlated dephasing therefore provides a more faithful effective model of bond-loss propagation and gives the decoder more reliable information.

Several refined approaches incorporate loss information into the decoding problem. Ref.~\cite{gu2024optimizingquantumerrorcorrection} tracks the possible time of loss relative to the gate order and passes this information to the decoder as modified edge weights. Other approaches incorporate loss configurations directly into the detector error model (DEM), allowing the decoder to condition on loss events~\cite{baranes2026leveraging}. This can capture correlated Pauli errors induced by missing bonds, but the effect of loss on the DEM is generally nonlinear: it cannot be represented by simply XOR-composing detector sets, as is possible for independent Pauli noise. Moreover, when later entangling gates involving the lost qubit are physically omitted rather than replaced by a Pauli channel, the resulting circuit may involve gauge-like degrees of freedom whose treatment in \texttt{STIM} is subtle~\cite{gidney_detector_stim_2024,gidney_nondeterministic_detectors_stim_2026}. The Pauli-envelope formalism offers another route by associating each loss event with the set of compatible Pauli configurations~\cite{liu2026achieving}. This restores linearity at the level of the detector description, but matching-based decoders must project the envelope onto an effective graph, which can degrade performance and reduce the effective distance.

Following Ref.~\cite{gu2024optimizingquantumerrorcorrection}, our approach is designed to retain the useful linear structure of a Pauli DEM while capturing the correlations induced by loss. It relies on two architectural features. First, photon loss is heralded at measurement. Second, each photon interacts with only a bounded number of atoms before being measured. We therefore partition each photon trajectory into intervals separated by its scheduled CZ gates. Conditioned on the photon being detected as lost, each interval is assigned the conditional probability that the photon was lost during that part of its trajectory. If the loss is assigned to a given interval, all subsequent CZ gates involving that photon are replaced by a correlated Pauli $Z$ error acting on the atoms that would have interacted with it. This represents the fact that, after loss, the photon can be treated as having undergone an unobserved projection, so the remaining scheduled interactions are represented by correlated $Z$-type uncertainty on the neighboring atoms, consistent with Refs.~\cite{yu2025processingdecodingrydbergdecay,yu2026locating}.

More explicitly, suppose that after loss in interval $k$, the photon would have interacted with atoms in the set $S_k$. The remaining scheduled interactions are represented by
$U_{S_k}=\prod_{a\in S_k} CZ_{p,a}$. Since 
$CZ_{p,a}=|0\rangle\langle0|_{p}\otimes I_a+|1\rangle\langle1|_{p}\otimes Z_a$, 
an unobserved projection of the photon leaves an unknown classical value $b\in\{0,1\}$, which applies $(Z_{S_k})^{b}$ to the atoms, where $Z_{S_k}=\prod_{a\in S_k}Z_a$. Averaging over the unknown value gives the correlated dephasing channel
$
\mathcal{E}_{S_k}(\rho)=\frac{1}{2}\rho+\frac{1}{2}Z_{S_k}\rho Z_{S_k}$,
which is a correlated $Z$ error channel on the atoms in $S_k$.

\begin{figure*}
    \centering
    \includegraphics[width=\linewidth]{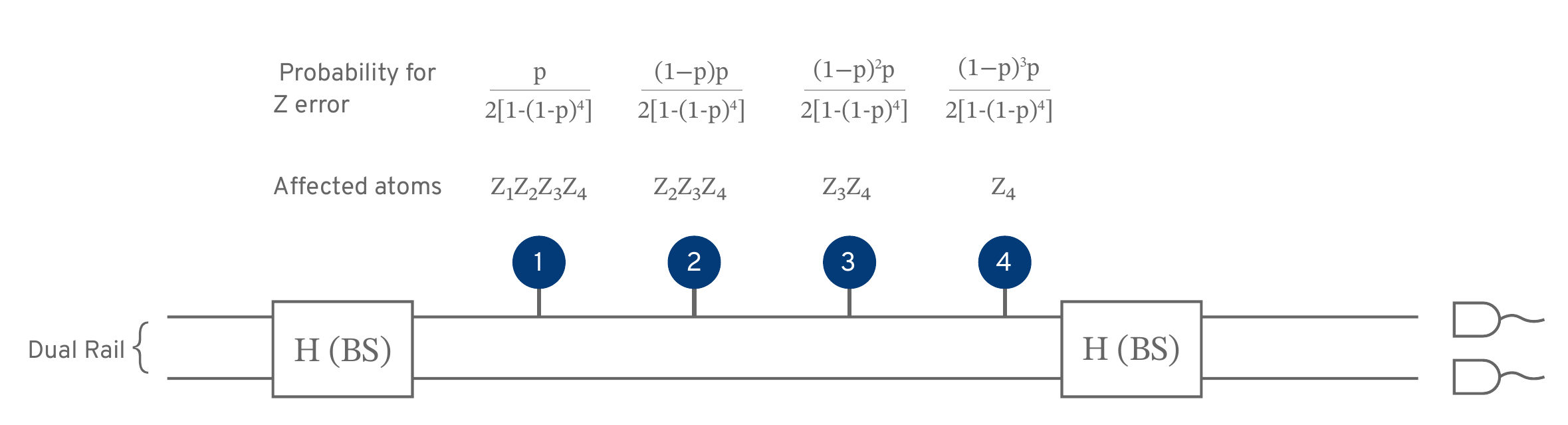}
    \caption{A dual-rail photonic qubit interacts sequentially with four atoms, shown as blue circles. The interactions implement CZ gates in the indicated order. The table lists the conditional probabilities that the photon was lost at each interaction, modeled as loss occurring immediately before that interaction, conditioned on the photon being found lost at measurement. It also shows the corresponding correlated $Z$ errors induced on atoms involved in subsequent interactions. The additional factor of $2$ in the probabilities arises because the photon is initialized in the $X$ basis and therefore occupies the $\ket{1}$ rail with probability $1/2$.}
    \label{fig:CondProbCZ}
\end{figure*}

An example is shown in Fig.~\ref{fig:CondProbCZ}. A single dual-rail photon, initialized in $\ket{+}$, undergoes four scheduled CZ gates with different atoms. Since bond-loss propagation affects only qubits that interact with the lost photon, the photon trajectory is divided into sections between these interactions. Each section is assigned the conditional probability that the photon was lost there, conditioned on the fact that the photon is absent at measurement. The table in Fig.~\ref{fig:CondProbCZ} lists the corresponding correlated $Z$ errors induced on the atoms involved in subsequent interactions. For clarity, the figure assumes that loss occurs only during entangling gates, but the construction extends directly to additional loss mechanisms. For example, if initialization loss occurs with probability $p_I$, the first interval probability is modified to $p_I + p(1-p_I)$, and the remaining conditional probabilities are renormalized.

We implement this loss model using a two-level Monte Carlo procedure. First, we sample $N$ heralded photon-loss configurations according to the physical loss model. Each sampled configuration specifies which photons are found to be lost at measurement. For each configuration, we construct the resulting DEM and run $M$ Monte Carlo shots. In each shot, the simulator samples one of the mutually exclusive candidate loss locations along each lost photon trajectory, together with the corresponding correlated Pauli error on the atoms affected by subsequent missing CZ gates.

The mutually exclusive candidate locations are encoded in \texttt{STIM} using \texttt{CORRELATED ERROR} and \texttt{ELSE CORRELATED ERROR} instructions~\cite{gidney2021stim}. This produces a ladder of conditional events for each lost photon, ensuring that exactly one loss interval is selected in a given realization. The probabilities in the ladder are chosen so that the cumulative distribution reproduces the desired conditional probabilities for the loss location. In this way, \texttt{STIM} samples a single consistent loss history for each lost photon while propagating the corresponding correlated Pauli errors to the detector record.

\begin{figure}
    \centering
    \includegraphics[width=0.95\linewidth]{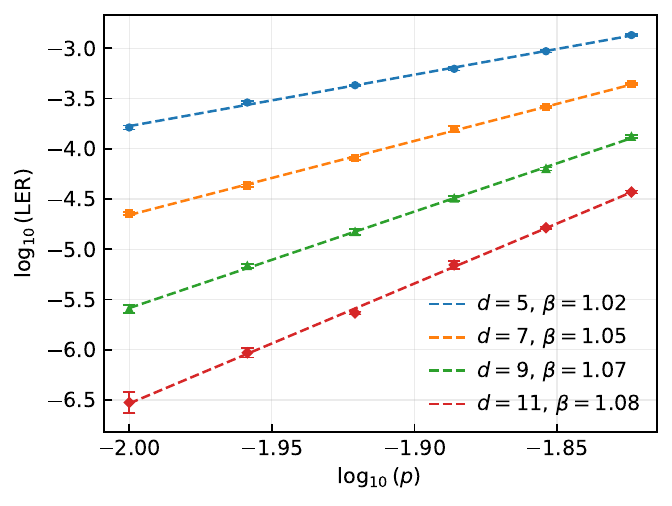}
    \caption{Below-threshold scaling of the logical error rate for the RHG memory-channel simulation under the circuit-level loss model of Table~\ref{tab:error_model}. The numerical data are fit to the expected form $p_L(p)=p_L(p_{\mathrm{th}})(p/p_{\mathrm{th}})^{\beta d}$ for $p \ll p_{\mathrm{th}}$, where $p_L$ denotes the LER, $p_{\mathrm{th}}$ is the threshold, $d$ is the code distance, and $\beta$ is a fitting parameter. We obtain $\beta \approx 1$ for all simulated distances, confirming optimal distance scaling of the decoder. 
    Each point uses at least $50{,}000$ loss realizations and $10{,}000$ decoding iterations per realization. Error bars show the standard error of the mean.}
    \label{fig:FittingMemory}
\end{figure}

After this replacement, the model is expressed entirely in terms of Pauli error mechanisms and can be decoded using standard minimum-weight perfect matching. We use \texttt{STIM} to sample detector events and PyMatching~\cite{higgott2021pymatchingpythonpackagedecoding} to perform MWPM decoding. Although the procedure involves an outer sampling over loss configurations, the high symmetry of the underlying codes reduces the number of effectively distinct configurations. In practice, a few thousand sampled loss configurations are sufficient for convergence in our simulations, which we verify by estimating the logical error rate with $95\%$ confidence intervals. Importantly, even under the full circuit-level loss model, our loss-aware decoder achieves optimal distance scaling for the logical error rate. That is, below threshold, the logical error rate decreases with the code distance $d$, as confirmed numerically in Fig.~\ref{fig:FittingMemory}.

\section{Clifford set in RHG lattice}\label{App:CliffordGates}

The Hadamard gate is defined by

\begin{align*}
H =
\frac{1}{\sqrt{2}}
\begin{pmatrix}
1 & 1 \\
1 & -1
\end{pmatrix},
\end{align*}
and satisfies $HX = ZH$. A logical Hadamard must therefore exchange the logical $X_L$ and $Z_L$ operators, or equivalently their correlation surfaces.

As explained in Sec.~\ref{App:Foliation}, physical information propagated from one RHG layer to the next acquires a Hadamard due to the cluster-state teleportation rule. This propagation alone is not a logical Hadamard. The reason is that the information associated with a logical correlation surface is stored as a string of physical $Z$ operators on the next layer, rather than being mapped to the complementary logical correlation surface. As in the surface code, a true logical Hadamard therefore requires an additional exchange of the logical directions~\cite{horsman2012surface}.

Using the nonlocal connectivity of our architecture, we implement this exchange by rotating the CZ connectivity rather than moving the qubits. The construction is illustrated in Fig.~\ref{Hgate}. Starting from a dual layer, corresponding to logical $\ket{0}_{L}$ initialization, we insert another dual layer instead of the usual primal layer. The CZ connections to this second dual layer are rotated by $90^{\circ}$ relative to the standard RHG connectivity.

In the RHG lattice, $X_L$ is represented by a vertical string of physical $Z$ operators on data qubits in a dual layer, while the $Z_L$ correlation surface is a horizontal string of physical $X$ measurements in the dual layer. After the $90^{\circ}$ rotation, the horizontal correlation surface in the first dual layer determines the vertical $Z$ string, and hence $X_L$, in the second dual layer. Conversely, the vertical $Z$ string representing $X_L$ in the first dual layer determines a horizontal $X$ correlation surface representing $Z_L$ in the second. Thus, the rotated connectivity exchanges the roles of $X_L$ and $Z_L$.

The rotation also shifts the geometric placement of the logical representatives. For example, the upper row representing $Z_{L}$ in the first dual layer determines the left column representing $X_{L}$ in the second dual layer, while the left column representing $X_{L}$ in the first dual layer determines the \textit{bottom} row representing $Z_{L}$ in the second. Unlike in the surface code, different representatives of the same logical operator in the RHG lattice are related by nondeterministic, but known, single-qubit $X$-basis measurement outcomes~\cite{fowler2008topological}. The conventional placement of the logical operators can therefore be restored in software by multiplying the correlation surface with the appropriate measurement outcomes.

The checks remain well defined after the rotation. Each check still contains six single-qubit $X$ measurements, but its type is exchanged: primal checks become dual checks and vice versa. In the language of topological defects, the logical Hadamard is implemented by a domain wall~\cite{bombin2023logical}. Importantly, this construction requires no additional qubits or gates. It is implemented entirely by modifying the CZ-connectivity pattern, so its logical error rate is essentially that of the identity channel \cite{bombin2023logical}, as shown in Fig.~\ref{fig: CliffordResults}(a).

\subsection{S gate}\label{app:Sgate} 
The phase gate is a $\pi/2$ rotation about the $Z$ axis,
\begin{align*}
 \begin{pmatrix}
    1 & 0 \\
    0 & i
\end{pmatrix}.   
\end{align*}
Under conjugation, it maps $X$ to $Y = i X Z$. A logical $S$ gate must therefore implement $X_L \to Y_L = i X_L Z_L$ while preserving the stabilizer structure.

A simple approach is to inject a logical $Y_L \propto \ket{0}_{L} + i\ket{1}_{L}$ resource state and use gate teleportation. In the RHG lattice, such a state can be initialized by a single-qubit $Y$-basis measurement~\cite{herr2018lattice}. This procedure is not fault tolerant, since the $Y$-basis measurement is not protected by checks, and therefore requires further distillation. Instead, we use a fold-transversal construction, following surface-code proposals based on diagonal reflection symmetry~\cite{moussa2016transversal, wang2023surface, breuckmann2024fold, webster2023transversal, chen2026transversal}. In this approach, the code is reflected along its diagonal mirror symmetry, and corresponding pairs of qubits on opposite sides of the fold are entangled, while physical $S$ and $S^{\dagger}$ gates are applied to the qubits lying along the diagonal. 

In the RHG formulation, the fold-transversal $S$ gate is implemented on a primal layer. As shown in Fig.~\ref{Str}, data qubits are connected by CZ gates to their mirror partners across the diagonal, while qubits on the diagonal are measured in alternating $Y^{+}$ and $Y^{-}$ bases, corresponding to physical $S$ and $S^{\dagger}$ operations. To see the action of the gate, consider the original $X_L$ correlation surface, which may be written as $X_1X_2X_3$. The diagonal operation maps $X_1$ to $Y_1$, while the additional CZ gates, for example $CZ_{2,4}$ and $CZ_{3,5}$, extend the correlation surface by adding $Z_4Z_5$. The resulting deterministic operator is therefore $Y_{1}X_{2}X_{3}Z_{4}Z_{5}$, which represents $Y_L = X_L Z_L$, since $Z_L$ is the horizontal logical $Z$ string on the primal layer.   

The stabilizer checks must be modified consistently with this transformation. For example, consider a check that before the gate contains $C = X_{1}X_{4}X_{a}$. After the fold-transversal operation, the check transforms to $\tilde{C} = (iX_{1}Z_{1})(-iX_{a}Z_{a})X_{4}Z_{2}$. Multiplying $\tilde{C}$ by the stabilizer $X_{b}Z_{1}Z_{a}Z_{2}$ centered at qubit $b$ cancels the extra $Z$ factors and restores an $X$-only check, now extended by the measured qubit $X_b$. The alternating pattern of $S$ and $S^{\dagger}$ on the diagonal ensures that the phase factors from $Y=iXZ$ cancel within each check, preserving the stabilizer structure and fault tolerance~\cite{wang2023surface}. 

This implementation has two main costs. First, each data qubit in the gate layer participates in one additional CZ gate. This introduces extra loss, but only within a single layer rather than throughout the bulk. Second, the modified checks produce hyperedges in the syndrome graph, so the detector error model is not directly matchable. As discussed in Appendix~\ref{App:Sim_details}, these hyperedges can be decomposed into matching-compatible subgraphs~\cite{cain2025fast, turner2025scalable}. Consequently, the logical $S$ gate remains efficiently decodable with MWPM and achieves a threshold comparable to the identity channel, with only a modest increase in logical error rate, as shown in Fig.~\ref{fig: CliffordResults}(b).

The main hardware-specific complication is that the diagonal CZ gates couple qubits of the same type, which conflicts with the bipartite photon--atom interaction pattern of the native architecture. This complication is confined to the gate layer.

Same-type effective CZ operations can be mediated using the opposite qubit type. For two atoms, a mediating photon can sequentially interact with both atoms, with an intermediate Hadamard on the photon, thereby producing the required effective entangling operation. Conversely, a mediating atom can generate the corresponding effective operation between photons. These mediated links are lossier than native photon--atom CZ gates, but they occur only in the single layer where the $S$ gate is implemented.

Previous analyses of confined layers of extra-lossy entangling gates indicate that such layers can tolerate substantially larger loss, by factors of order $5$--$10$, while producing only modest threshold or LER degradation~\cite{ramette_fault-tolerant_2024, stack2026transversalfaulttolerantdistributed}. The additional loss can also be reduced by using a STAP-based implementation: the upper triangular part of the gate layer is initialized as atoms and the rest as photons, the required diagonal CZ gates are performed, and STAP is then applied to the atoms. Motivated by these considerations, our simulations use a simplified model in which same-type diagonal links are allowed directly in the relevant layer and are assigned the same loss probability as native photon--atom links. A full mediated-link simulation is left for future work.

The fold-transversal construction is substantially more resource efficient than local-connectivity alternatives. In the logical-block construction of Ref.~\cite{bombin2023logical}, the rotated-planar-code phase gate has spatial footprint $2d^2$ for $2d$ syndrome-measurement rounds, giving space-time volume $4d^3$. A more sophisticated in-place approach due to Gidney~\cite{gidney2024inplace} avoids expanding the patch bounding box, but implements the operation through diagonal twist motion and modified syndrome-measurement rounds, with a $Y$-basis initialization or measurement primitive requiring $d/2+O(1)$ syndrome-measurement rounds. By contrast, with nonlocal connectivity the RHG fold-transversal $S$ gate requires only one additional CZ gate per data qubit in a single layer, with no additional logical qubits and no additional time steps.

\subsection{CNOT gate} \label{CNOT}

The only two-qubit gate required to complete the Clifford set is CNOT. It is defined by
\begin{align*}
   \begin{pmatrix}
    1 & 0 & 0 & 0 \\
    0 & 1 & 0 & 0 \\
    0 & 0 & 0 & 1 \\
    0 & 0 & 1 & 0 
\end{pmatrix}. 
\end{align*}
In surface-code and RHG architectures, the two standard fault-tolerant implementations are lattice surgery, which uses merge-and-split operations through joint stabilizer measurements~\cite{horsman2012surface, herr2018lattice, chamberland2022universal}, and transversal CNOT, which uses pairwise interactions between physical qubits of the two logical blocks.

We use the nonlocal connectivity of the compound architecture to implement the transversal version. Pairwise CZ gates are applied between the primal layer of the control logical block and the dual layer of the target logical block. This implements the required logical transformations $X^{c}_{L} \to X^{c}_{L}X^{t}_{L}$ and $Z^{t}_{L} \to Z^{c}_{L}Z^{t}_{L}$ while leaving $Z_L^c$ and $X_L^t$ invariant. One of these transformations, $I^c Z_L^t \to Z_L^c Z_L^t$, is illustrated in Fig.~\ref{CNOTgate} 

The checks on the affected layers must be redefined consistently with the transversal CZ layer. For example, consider data qubits on the primal layer of the control block that participate in the same check. Because each interacts transversally with the target block, their stabilizers acquire additional $Z$ factors. These extra contributions are canceled by incorporating the appropriate SL qubit from the dual layer of the target block, preserving the check and making the logical transformation fault tolerant. A logical CZ can be implemented similarly by connecting the two logical blocks through the rotated dual-layer geometry used for the Hadamard construction in Fig.~\ref{Hgate}.

Compared with lattice surgery, the transversal construction has much smaller space-time overhead. Lattice-surgery CNOT implementations require an additional logical ancilla patch of size $O(d^2)$ and $O(d)$ syndrome-measurement rounds for each joint logical measurement, such as $XX$ and $ZZ$, although some optimized circuits require only one of these measurements~\cite{akahoshi2025runtime}. By contrast, the transversal CNOT uses a single layer of pairwise CZ gates between the two existing logical blocks. It requires no additional logical block and only $O(1)$ temporal overhead.

As with the fold-transversal $S$ gate, the transversal CNOT introduces an additional layer of CZ gates and relies on nonlocal connectivity. Unlike the $S$ gate, however, it is compatible with the bipartite structure of the native hardware, because the added CZ gates couple qubits of opposite type. The resulting detector error model still contains hyperedges. As discussed in Appendix~\ref{App:Sim_details}, these can be decomposed into matching-compatible subsets, allowing the decoding problem to be treated with MWPM~\cite{cain2025fast, turner2025scalable}. As shown in Fig.~\ref{fig: CliffordResults}(c), this preserves the memory-channel threshold, but increases the LER.

\bibliography{qs_arch_bp_clean}

\end{document}